\documentclass[preprint,prd,showpacs,nofootinbib,showlabels]{revtex4}

\usepackage{graphicx}
\usepackage{amsmath,amsfonts,amssymb}
\usepackage{array}
\usepackage[pdftex]{color}
\usepackage[sort&compress]{natbib}
\usepackage[colorlinks=true,linkcolor=blue,filecolor=blue,urlcolor=blue,citecolor=blue,pdftex=true,plainpages=false]{hyperref}
\usepackage[mathscr]{euscript}

\def\negcdot{\negmedspace\cdot\negmedspace}

\newcommand*{\chpt}{\raise0.4ex\hbox{$\chi$}PT}
\newcommand*{\schpt}{S\raise0.4ex\hbox{$\chi$}PT}
\def\rschpt{rS\raise0.4ex\hbox{$\chi$}PT}
\def\rhmschpt{HMrS\raise0.4ex\hbox{$\chi$}PT}
\def\aschpt{HMrAS\raise0.4ex\hbox{$\chi$}PT}

\newcommand*{\ie}{\textit{i.e.},\ }
\newcommand*{\eg}{\textit{e.g.},\ }
\newcommand*{\vs}{\textit{vs.}\ }
\newcommand*{\etc}{\textit{etc.}}

\providecommand*{\prd}[1]{Phys.\ Rev.\ \textbf{D#1}}
\renewcommand*{\prd}[1]{Phys.\ Rev.\ \textbf{D#1}}

\newcommand*{\MeV}{{\rm Me\!V}}

\newcommand{\opopo}{\ensuremath{1\!+\!1\!+\!1}}

\newcommand*{\Tr}{\ensuremath{\operatorname{Tr}}}

\newcommand{\trD}{\ensuremath{\textrm{tr}_{\textrm{\tiny \it D}}}}

\newcommand{\trDt}{\ensuremath{\textrm{Tr}}}

\def\gtwid{{\,\raise.35ex\hbox{$>$\kern-.75em\lower1ex\hbox{$\sim$}}\,}}
\def\ltwid{{\,\raise.35ex\hbox{$<$\kern-.75em\lower1ex\hbox{$\sim$}}\,}}
\def\leftvec{{\raise1.5ex\hbox{$\leftarrow$}\kern-1.00em}}
\def\rightvec{{\raise1.5ex\hbox{$\rightarrow$}\kern-1.00em}}
\def\half{{\scriptstyle \raise.2ex\hbox{${1\over2}$}}}
\def\threehalves{{\scriptstyle \raise.15ex\hbox{${3\over2}$}}}
\def\third{{\scriptstyle \raise.15ex\hbox{${1\over3}$}}}
\def\third{{\scriptstyle \raise.15ex\hbox{${1\over3}$}}}
\def\twothirds{{\scriptstyle \raise.15ex\hbox{${2\over3}$}}}
\def\fourth{{\scriptstyle \raise.15ex\hbox{${1\over4}$}}}

\newcommand{\Dslash}{\ensuremath{D\!\!\!\! /}}
\newcommand{\Bslash}{\ensuremath{B\!\!\!\! /}}

\newcommand{\vslash}{\ensuremath{v\!\!\! /}}

\newcommand{\cL}{\ensuremath{\mathcal{L}}}
\newcommand{\cM}{\ensuremath{\mathcal{M}}}

\newcommand{\cO}{\ensuremath{\mathcal{O}}}

\newcommand{\cV}{\ensuremath{\mathcal{V}}}

\def\eqn#1{\label{eq:#1}}
\newcommand{\eq}[1]{Eq.~\eqref{eq:#1}}

\newcommand{\eqs}[2]{Eqs.~\eqref{eq:#1} and \eqref{eq:#2}}

\newcommand{\eqsthru}[2] {Eqs.~\eqref{eq:#1} through \eqref{eq:#2}}

\newcommand{\eqsthree}[3]{Eqs.~\eqref{eq:#1}, \eqref{eq:#2} and \eqref{eq:#3}}

\def\figref#1{Fig.~\ref{fig:#1}}

\def\secref#1{Sec.~\ref{sec:#1}}

\def\Secref#1{Section~\ref{sec:#1}}
\def\tabref#1{Table~\ref{tab:#1}}
\def\rcite#1{Ref.~\cite{#1}}
\def\rcites#1{Refs.~\cite{#1}}

\begin{document}

\title{Chiral Perturbation Theory\\ for All-Staggered Heavy-Light Mesons}
\author{Claude Bernard and Javad Komijani}
\affiliation{Washington University, St.\ Louis, MO 63130}

\begin{abstract}
In highly improved staggered quark (HISQ) simulations by the HPQCD, MILC, and
Fermilab Lattice collaborations, both the light quarks and the charm quark are staggered.
We extend chiral perturbation theory for staggered quarks to 
include such all-staggered heavy-light mesons.
We assume that the heavy quark action is sufficiently improved that we may take $a m_Q <<1$
(where $m_Q$ is the heavy quark mass),
but also that $m_Q>>\Lambda_{QCD}$ so that a continuum heavy quark expansion is appropriate.
We develop this effective chiral theory through next-to-leading order,
and use it to study the pattern of taste splittings in the heavy-light meson
and to compute the leptonic decay constant of the heavy-light
meson to one-loop in the chiral expansion. 
\end{abstract}

\maketitle

\section{Introduction}
\label{sec:Intr}

Heavy-light meson systems provide some of the best ways to test
the standard model and look for signs of new physics.  In particular, the
constraints on the sides of the unitarity triangle, which
come mainly from heavy-light decays and mixings, are limited largely by
the size of the theoretical errors in the values of the hadronic matrix elements
of weak operators.  Lattice QCD provides a means of carrying out 
non-perturbative calculations of such quantities from first principles and with
controlled errors.  

In setting up a lattice QCD calculation, a key choice is the form of the lattice action
for the quarks.
Staggered fermions \cite{Kogut:1974ag} are an efficient approach to simulating light quarks.
The ``highly improved staggered quark'' (HISQ) action \cite{HISQ} makes it possible 
to treat charm quarks with the same action as the light quarks.  Thus 
``all-staggered'' simulations of $D$ and $D_s$ mesons are now possible \cite{HISQ-charm,Bazavov:2012dg}, 
and even $B_s$ mesons have been treated in this way by pushing up the heavy quark mass 
on ensembles with the finest available lattice spacings \cite{McNeile:2011ng}. 

There are several advantages to this all-staggered approach.  Since heavy and light
quarks have the same action, there are partially conserved heavy-light axial and vector
currents that need no renormalization.  The tuning of the
heavy quark mass is also simplified compared to other approaches 
(see, for example, Ref.~\cite{Bernard:2010fr}) because difference between ``rest'' and
``kinetic'' masses of the heavy quark due to discretization effects may be neglected.
Further, the statistical errors of heavy-light pseudoscalars tend to be rather small,
as they are for light-light staggered pseudoscalars.

Lattice computations often involve an extrapolation in light
quark masses to the physical up and down masses, and always require a continuum
extrapolation in lattice spacing.  A version
of chiral perturbation theory (\chpt) that includes the effects of the
discretization errors can help to control these extrapolations. Here, we develop
chiral perturbation theory for all-staggered heavy-light mesons.   We call
the theory {\it heavy-meson, rooted, all-staggered chiral perturbation theory} (\aschpt),
where ``rooted'' refers to the fourth root of the staggered determinant, as reviewed below.

Staggered quarks have a four-fold degree of freedom, called taste, which is a remnant
of lattice doubling. In the continuum limit, there is an exact $SU(4)$ symmetry acting
on tastes; this symmetry is broken at $\cO(a^2)$ in the lattice spacing $a$. 
The corresponding discretization errors in the light-light sector split the masses of
mesons with different tastes, which may be 
understood using staggered chiral perturbation theory (\schpt) \cite{LEE_SHARPE,SCHPT}.
For typical values of $a^2$, 
the taste splittings of light pseudoscalar
mesons can be comparable to the masses themselves.  In short-hand, we say
$a^2\sim m_\pi^2$,  where factors of $\Lambda_{QCD}$ to balance the dimensions are always assumed
in such relations.  These taste splittings must therefore be included in the leading order (LO) light-light Lagrangian.   

For heavy-light mesons composed of staggered quarks,
the situation is different. The LO Lagrangian in the continuum is of ${\cO}(k)$,
where $k$ is the residual momentum of the heavy-light meson. We
assume $k\sim m_\pi$.  Since  $a^2\sim m_\pi^2\sim k^2$,
taste violations are of higher order and will be treated 
as next-to-leading order (NLO) corrections.
The LO heavy-light Lagrangian is then taste invariant.
This power counting is consistent with HISQ simulations, where the splittings in {\em squared}\/ meson
masses remain roughly constant as the valence quark mass increases from the light
quark regime to the charm regime \cite{Bazavov:2012xda}. Therefore the splittings for the masses themselves
are much smaller for heavy-light mesons than for light mesons.
For example, the 
taste splitting at $a\approx0.12\;$fm
between the root-mean-squared (RMS) $D_s$ meson and the lightest
$D_s$ meson is only about 11 MeV \cite{Bazavov:2012xda}, while it is
about 110 MeV for the pion.

Reference \cite{Aubin:StagHL2006} works out a closely related chiral theory 
for heavy-light mesons with
staggered light quarks but non-staggered heavy quarks (for example,
Fermilab \cite{El-Khadra:1996mp} or NRQCD \cite{NRQCD} quarks). That chiral theory
has  been called {\it heavy-meson, rooted staggered chiral perturbation theory} (\rhmschpt).  
In \rhmschpt, heavy-light
mesons have a single taste degree of freedom associated with the light quark.
As in the current case, the LO \rhmschpt\ Lagrangian
in the heavy-light sector is taste invariant.%
\footnote{There is in fact
is no mass splitting of different tastes of heavy-light mesons at any order in \rhmschpt.
The absence of splittings is guaranteed by 
shift symmetry \cite{SHIFT,BGS08}, which in the continuum limit is simply a discrete subgroup
of continuum $SU(4)$ taste symmetry.}  
Since the LO Lagrangian determines
the propagators and vertices of the one-loop diagrams, those diagrams 
are very closely related in \rhmschpt\ and \aschpt\ (the current case). Important
differences arise at NLO, however.  Such differences affect, for example, the analytic
terms that are added on to the one-loop chiral logarithms to give the complete NLO
expressions for quantities such as the decay constants.  Similarly, mass splittings for heavy-light
mesons of different tastes are governed by the analytic NLO terms.  Indeed, we
prove below that the one-loop diagrams themselves  do not give rise to any taste violations in
the heavy-light meson masses, despite the fact the light-light masses, which enter those diagrams,
do violate taste symmetry.  This feature arises from the combination of exact heavy-quark taste symmetry at LO
and the all-orders discrete taste symmetry coming from shift invariance.

Thus we need to extend the program developed in Ref.~\cite{Aubin:StagHL2006} 
to include staggered heavy quarks with a taste degree of freedom. 
In this paper we assume that the staggered action used (\eg HISQ) 
is improved sufficiently that
we can treat the heavy quark as ``continuum-like,'' with small corrections
from cutoff effects.  We refer to this assumption in short-hand as taking  $am_Q\ll1$, 
where $m_Q$ is mass of the heavy quark, although one should keep in mind that corrections
in powers of $am_Q$ may in practice be reduced as much or more by the improved action than by
the size of $am_Q$ {\em per se}. 
Under this assumption, we can use the Symanzik Effective Theory (SET)  \cite{SET} to describe
the discretization effects on the heavy quarks, as well as on the light quarks.
The SET is the
effective theory for physical momenta $p$ small compared
with the cutoff ($ap\ll1$); it  encodes discretization
effects  in higher-dimensional operators added to continuum QCD.

When the heavy quark
is non-staggered, as in
\rhmschpt, the heavy-quark doubler states 
are split from the heavy quark by an amount of order of the cutoff, and are therefore
integrated out of the SET. Thus the heavy quark fields have no degree of freedom 
corresponding to taste, and
taste violations at $\cO(a^2)$ appear only in
four-quark operators composed exclusively of light quarks.

In the all-staggered case, on the other hand, important taste violations at  $\cO(a^2)$ 
appear in ``mixed'' four-quark operators consisting of the product of 
a heavy quark bilinear and a light quark bilinear, as well as in the product of two 
light-quark bilinears.  These operators break the taste symmetries of both
heavy and light quarks.  (Products of
two heavy-quark bilinears also appear in the SET, but their effect on the heavy-light
meson Lagrangian is rather trivial since there is at most one heavy quark in all initial
and final states considered.)

In the SET, the lattice theory has been replaced by a continuum theory.  The lattice spacing
$a$ appears only as a parameter multiplying higher-dimensional operators. One can
then use the fact that 
$m_Q$ is large compared to $\Lambda_{QCD}$, to organize heavy quark effects
with  Heavy Quark Effective Theory (HQET).  The heavy quark field $q_h$ in both 
dimension-four  and higher-dimension operators is replaced by a 
HQET field $Q$, where $Q$ satisfies
\begin{equation}
\eqn{Qdef}
\frac{1+\vslash}{2}\; Q = Q \ ,
\end{equation}
with $v_\mu$ the heavy-quark four-velocity.
The dimension-four terms are invariant under
heavy-quark spin symmetry, but the higher dimensional terms may violate the symmetry.

Finally, when residual momenta and light quark masses are small compared to the chiral
scale $\Lambda_\chi\sim 1\;$GeV, the physics of light-light and heavy-light mesons
may be described by a chiral effective theory.
The dimension-four operators give a standard-looking heavy-meson chiral theory, but
with additional taste degrees of freedom for both light and heavy quarks.
The higher-dimensional operators may be mapped to the chiral Lagrangian
using a spurion analysis.  They generate LO terms in the light-light sector that violate
light-quark taste symmetry, and NLO terms in the heavy-light sector that violate heavy-quark
taste and spin symmetry.

Since the four taste degrees of freedom of a staggered quark are unphysical, 
the fermion determinant is replaced by its fourth root in simulations.
This rooting procedure introduces non-locality:
At non-zero lattice spacing, the rooted fermion action is not equivalent to any 
local action \cite{BGS06},  which in turn leads to nonlocal violations of unitarity
\cite{Prelovsek05,BGS06}. In the continuum limit, locality and unitarity are
however expected to be restored, an expectation which is supported
theoretical arguments~\cite{Shamir04,Shamir06,Bernard06,BGS08},
as well as other analytical and numerical
evidence~\cite{SharpePoS06,KronfeldPoS07,GoltermanPoS08,Bazavov:2009bb,Donaldetal11}.

In the chiral theory, rooting is taken into account
by multiplying each sea quark  loop by a factor of $1/4$
\cite{Aubin:2003mg,Aubin:2003uc}.  This can be accomplished  either
by following the quark flow \cite{QUARK-FLOW} to locate the loops,
or --- more systematically --- by replicating the sea quarks $n_r$, performing
a standard chiral calculation, and
taking $n_r=1/4$ in the result\cite{Bernard06,BGS08}.
Here, we follow Ref.~\cite{Aubin:StagHL2006} and use the quark flow approach.

After the chiral theory is constructed, we first apply it to calculate 
the  taste splittings of heavy-light meson masses at next-to-leading chiral order.    
Some of the analytic NLO terms break the taste-$SU(4)$ symmetry of the masses 
down to  $SO(4)$ symmetry \cite{LEE_SHARPE}, while others break the symmetry still further, producing
splitting within $SO(4)$ multiplets.   Our results can be used to understand the measured lattice splittings
\cite{Bazavov:2012xda}.  

We then calculate
the leptonic decay constant of a heavy-light meson at one-loop. 
The chiral form we obtain is very useful in the analysis of HISQ data for $f_{D^+}$ and $f_{D_s}$
\cite{work-in-progress}. 
In general, we work to LO in $1/m_Q$, but some higher order terms (heavy-light
hyperfine and flavor splittings) are considered in
the  decay constant calculation. Following Ref.~\cite{FermilabMILC_Dec2011},
we argue that the inclusion of those terms
(but no other $1/m_Q$ terms) constitutes a systematic approximation in 
the power counting introduced by Boyd and Grinstein \cite{BoydGrinstein}.

As is clear from the above, many features  of the analysis
of \rcite{Aubin:StagHL2006} can be used here with only small changes.  However,
in reexamining the NLO terms in the Lagrangian and current of \rcite{Aubin:StagHL2006}
for use here, we have discovered some minor mistakes: There are a few 
terms at NLO that were omitted, and a few of the terms listed in the earlier paper 
can be shown either to be absent or to be redundant with terms already present. 
This occurs only for the complicated
terms that violate both (Euclidean) rotation symmetry and taste symmetry.  
The errors have no consequences for applications of \rhmschpt\ in the literature.

The remainder of this paper is organized as follows: 
In \secref{ls-lag}, the LO \schpt\ Lagrangian is constructed 
for all-staggered heavy-light mesons, 
and those NLO terms that are the same as in the continuum are briefly discussed.
The $\cO(a^2)$ terms involving heavy-light mesons are then derived from a spurion analysis in
\secref{4quark-ops}, with a needed reduction of a three-index Lorentz tensor into irreducible
representations relegated to Appendix \ref{sec:tensor}.
\Secref{Mass} focuses on taste splittings of heavy-light mesons. 
Finally, in \secref{fD}, the decay constant in heavy-light systems is calculated to NLO.
Our conclusions and some discussion of the results follow in \secref{conclusions}.

\section{The staggered chiral Lagrangian with heavy-light mesons}
\label{sec:ls-lag}

In this section,  we first introduce our chiral power counting and give our notation for
the various contributions that appear at both LO and NLO.
We then consider the LO Lagrangian for both the light mesons and heavy-light mesons.
The heavy-light meson field is generalized from that in \rcite{Aubin:StagHL2006}  so that it carries a heavy-quark
taste index, in addition to light-quark taste and flavor --- or, equivalently, so that it carries meson taste and light-quark
flavor indices.  The NLO terms that are invariant under taste symmetry are the same as in
the continuum, and are briefly  treated in \secref{NLO-CONTINUUM}.

\subsection{Power counting}
\label{sec:power}

We assume the power counting $p_\pi^2\sim m^2_\pi\sim m_q\sim a^2$ for the light mesons (``pions'')
as in Ref. \cite{Aubin:StagHL2006}. Here $p_\pi$ is a typical pion momentum, and factors of $\Lambda_{QCD}$
are implicit. Two additional scales enter with the inclusion of heavy-light mesons. The 
first is the residual momentum of the heavy-light meson, $k$, which we take to be of the same order as $p_\pi$. 
The second scale is the heavy quark mass $m_Q$.  Initially, we keep only the leading order in $1/m_Q$ 
in the following calculations and derive the decay constant of $D$ at that order. We then
follow \rcite{FermilabMILC_Dec2011} to include hyperfine splittings
(e.g., $m^*_D-m_D$) and flavor splittings (e.g., $m_{D_s}-m_D$) in the NLO decay constant calculation. 
These splittings are $\sim\! 100$ \MeV, and so not much smaller than $m_\pi$, despite the fact that
they are formally of order $1/m_Q$.  Including the splittings can therefore be important
in practical applications of our results, especially since HISQ simulations at physical
pion mass are now available \cite{Bazavov:2012xda}.
 Furthermore it  is consistent
to include the splittings 
at NLO in the power counting of \rcites{BoydGrinstein,FermilabMILC_Dec2011}, 

The LO chiral Lagrangian is therefore $\cO(k\!\sim\!\! \sqrt{m_q})$ in the heavy-meson fields
and $\cO(m_q,a^2)$ in the light-meson fields. (As usual in HQET,
terms of $\cO(k^0)$ in the heavy-meson fields, \ie heavy 
mass terms,  are removed by construction.) 
Since each loop will bring in two powers
of $p_\pi$ or equivalent scales, we consider terms both of 
order $k^2$ and of order $k^3$
in the heavy mesons to be NLO, and similarly 
next-to-next-to-leading order (NNLO)
would include heavy-meson terms of order $k^4$ and $k^5$. For our purposes
here, we need the complete LO Lagrangian (for both heavy and light mesons), but only
the heavy-meson part of the NLO Lagrangian. We therefore write
\begin{eqnarray}
\cL           & = & \cL_{\rm LO} + \cL_{\rm NLO}\ ,\label{eq:Lcomplete}\\
 \cL_{\rm LO} & = & \cL_{\rm LO}^{\rm pion}+ \cL_1\ , \label{eq:L_LO}\\
\cL_{\rm NLO} & = & \cL_2 + \cL_3 \, \label{eq:L_NLO}
\end{eqnarray}
where $\cL_{\rm LO}^{\rm pion}$ is the standard LO light meson Lagrangian \cite{SCHPT}, 
and $\cL_1$, $\cL_2$, and $\cL_3$ denote the heavy-meson terms of order $k^1$, $k^2$ and
$k^3$ (or equivalent scales), respectively. 

We will also need $j^{\mu,i\Xi}$, the left-handed heavy-light current for light flavor $i$
and combined taste $\Xi$.  It has the similar expansion
\begin{eqnarray}
j^{\mu,i \Xi} & = & j^{\mu,i \Xi}_{\rm LO} + j^{\mu,i \Xi}_{\rm NLO}\ ,\label{eq:completecurrent}\\
j^{\mu,i \Xi}_{\rm NLO} & = & j^{\mu,i \Xi}_{1} + j^{\mu,i \Xi}_2 \ , \label{eq:NLOcurrent}
\end{eqnarray}
where again the subscripts 1 and 2 denote orders in $k$.

We can classify contributions to 
the NLO terms in \eqs{L_NLO}{NLOcurrent} by the source of the
extra powers of the scale and the nature of any symmetry breaking.  
The subscript $k$ will denote terms in which the powers come
exclusively from additional derivatives as compared to the LO terms, while
the subscripts $m$ and $a^2$ will indicate insertions of mass or taste-violating
spurions, respectively (together with possible additional derivatives).  
The taste-violating terms may be further classified according to 
whether continuum Euclidean $SO(4)$ rotation symmetry is preserved  or
broken (``type A'' or ``type B,'' respectively), 
and whether the heavy-quark taste symmetry is preserved or broken 
(``type 1'' or ``type 2'', respectively). As first pointed out in \rcite{LEE_SHARPE},
type A terms also preserve a $SO(4)$ taste symmetry of the light quarks, and that
feature remains true here.
Our classification then gives
\begin{eqnarray}
\cL_2	      & = & \cL_{2,k} + \cL_{2,m} + \cL_{2,a^2}^{A1} +  \cL_{2,a^2}^{B1}
+ \cL_{2,a^2}^{A2} +  \cL_{2,a^2}^{B2} \ , \label{eq:L2}\\
\cL_3         & = & \cL_{3,k} + \cL_{3,m} + \cL_{3,a^2}^{A1} +  \cL_{3,a^2}^{B1}
+ \cL_{3,a^2}^{A2} +  \cL_{3,a^2}^{B2} \ , \label{eq:L3}
\\
j^{\mu,i \Xi}_{1} & = & j^{\mu,i \Xi}_{1,k} \ , \label{eq:current1}\\
j^{\mu,i \Xi}_2 & = & j^{\mu,i \Xi}_{2,k} + j^{\mu,i \Xi}_{2,m}\
           + j^{\mu,i \Xi}_{2,a^2,A1} +  j^{\mu,i \Xi}_{2,a^2,B1
}+ j^{\mu,i \Xi}_{2,a^2,A2} +  j^{\mu,i \Xi}_{2,a^2,B2} \ , \label{eq:current2} 
\end{eqnarray}
where $j^{\mu,i \Xi}_{1}$ comes solely from derivative terms, since mass and taste spurions
bring in two powers of the small scale.

After introducing our (mainly standard) notation, we give the LO terms
$\cL_{\rm LO}^{\rm pion}$, $\cL_1$, and $ j^{\mu,i \Xi}_{\rm LO} $ 
in the next subsection. NLO terms that are the same as in the continuum,
namely $\cL_{2,k}$, $\cL_{3,k}$, $\cL_{2,m}$, $\cL_{3,m}$,
$j^{\mu,i \Xi}_1 $, $j^{\mu,i \Xi}_{2,k}$, and $j^{\mu,i \Xi}_{2,m}$ are then briefly
discussed in \secref{NLO-CONTINUUM}. Study of the taste-violating terms, which
require a detailed look at the SET,  are postponed until \secref{4quark-ops}.
Those terms that preserve heavy-quark taste symmetry, namely type A1 and B1 terms, 
are trivial generalizations of the corresponding terms in \cite{Aubin:StagHL2006}. Those that break heavy-quark taste
symmetry, namely type A2 and B2, are however completely new.

\subsection{Leading-order theory}
\label{sec:LOcont}

The LO chiral Lagrangian is divided into the light meson part 
$\cL_{\rm LO}^{\rm pion}$ and the heavy meson part $\cL_1$, 
as in \eq{L_LO}.
The light meson part is standard \cite{SCHPT}. However,
following \rcite{Aubin:StagHL2006}, we write the complete
Lagrangian in Minkowski space for ease
of comparison with the continuum heavy-light literature.
If desired, a Wick rotation can be defined everywhere to
transform the theory into Euclidean space, corresponding to the Euclidean lattice theory.
We have
\begin{equation}\label{eq:Lpion}
\cL_{\rm LO}^{\rm pion} =  \frac{f^2}{8} \Tr(\partial_{\mu}\Sigma \partial^{\mu}\Sigma^{\dagger})
                         + \frac{1}{4}\mu f^2 \Tr(\cM\Sigma+\cM\Sigma^{\dagger})
                         - \frac{2m_0^2}{3}(U_I + D_I + S_I+\ldots)^2 
			  - a^2\cV_\Sigma,
\end{equation}
where $\Sigma = \exp [i\Phi/f]$ is a $4n \times 4n$ matrix for $n$ staggered flavors, with $\Phi$ given by:
\begin{eqnarray}\label{eq:Phi}
  \Phi = \left( \begin{array}{cccc}
      U  & \pi^+ & K^+ & \cdots \\*
      \pi^- & D & K^0  & \cdots \\*
      K^-  & \bar{K^0}  & S  & \cdots \\*
      \vdots & \vdots & \vdots & \ddots \end{array} \right).
\end{eqnarray}
Here $U = \sum_{\Xi=1}^{16} U_\Xi T_\Xi$, \etc, with the Hermitian taste
generators $T_\Xi$ given by
\begin{equation}\label{eq:T_Xi}
  T_\Xi = \{ \xi_5, i\xi_{\mu 5}, i\xi_{\mu\nu} , \xi_{\mu}, \xi_I\}\ .
\end{equation}
As in Ref. \cite{Aubin:StagHL2006}, we employ Euclidean gamma matrices for $\xi_\mu$, 
with $\xi_{\mu\nu}\equiv (1/2)[\xi_{\mu},\xi_{\nu}]$ ($\mu <\nu$ in \eq{T_Xi}), $\xi_{\mu 5}\equiv \xi_{\mu}\xi_{5}$, 
and $\xi_I\equiv I$, where $I$ is the $4\times 4$ identity matrix.
Below, we use a summation convention for indices on the matrices $\xi_\mu$
that are repeated twice, but
explicit summation for indices that are
repeated more than twice.
The mass matrix is given by the $4n\times 4n$ matrix
\begin{eqnarray}
  \cM = \left( \begin{array}{cccc}
      m_u I  & 0 &0  & \cdots \\*
      0  & m_d I & 0  & \cdots \\*
      0  & 0  & m_s I  & \cdots\\*
      \vdots & \vdots & \vdots & \ddots \end{array} \right).
\end{eqnarray}
The potential $\cV_\Sigma$, which breaks the taste symmetry of light mesons, is defined in 
Refs.~\cite{SCHPT,Aubin:StagHL2006}:
\begin{eqnarray}
        -\cV_\Sigma & = & C_1
         \Tr(\xi^{(n)}_5\Sigma\xi^{(n)}_5\Sigma^{\dagger})
+\frac{C_3}{2} [ \Tr(\xi^{(n)}_{\nu}\Sigma
        \xi^{(n)}_{\nu}\Sigma) + h.c.] \nonumber \\*
        & & +\frac{C_4}{2} [ \Tr(\xi^{(n)}_{\nu 5}\Sigma
        \xi^{(n)}_{5\nu}\Sigma) + h.c.]
+\frac{C_6}{2}\ \Tr(\xi^{(n)}_{\mu\nu}\Sigma
        \xi^{(n)}_{\nu\mu}\Sigma^{\dagger}) \nonumber \\*
        & & + \frac{C_{2V}}{4}
                [ \Tr(\xi^{(n)}_{\nu}\Sigma)
        \Tr(\xi^{(n)}_{\nu}\Sigma)  + h.c.]
        +\frac{C_{2A}}{4} [ \Tr(\xi^{(n)}_{\nu
         5}\Sigma)\Tr(\xi^{(n)}_{5\nu}\Sigma)  + h.c.] \nonumber \\*
        & & +\frac{C_{5V}}{2} [ \Tr(\xi^{(n)}_{\nu}\Sigma)
        \Tr(\xi^{(n)}_{\nu}\Sigma^{\dagger})]
         +\frac{C_{5A}}{2} [ \Tr(\xi^{(n)}_{\nu5}\Sigma)
        \Tr(\xi^{(n)}_{5\nu}\Sigma^{\dagger}) ]\ .
\eqn{V}
\end{eqnarray}
The explicit $4n\times 4n$ matrices $\xi^{(n)}_{\mu}$ in \eq{V} are defined by
\begin{equation}\label{eq:xi-n}
  \left(\xi^{(n)}_{\nu}\right)_{ij} = \xi_{\nu}\delta_{ij}\ ,
\end{equation}
with $i$ and $j$ the $SU(n)$ light quark flavor indices,
and $\xi_\nu$ a $4\times4$ taste matrix, as in \eq{T_Xi}.
The matrices $\xi^{(n)}_{\mu\nu}$ and $\xi^{(n)}_{\nu5}$ are defined similarly.

In terms involving heavy-lights, we also need $\sigma \equiv \sqrt{\Sigma} = \exp[ i\Phi / 2f ]$. Both $\Sigma$ and $\sigma$ are 
singlets under the heavy-quark symmetries, while under $SU(4n)_L\times SU(4n)_R$ they transform as
\begin{eqnarray}
  \Sigma \to  L\Sigma R^{\dagger}\,,\qquad&&\qquad
  \Sigma^\dagger \to  R\Sigma^\dagger L^{\dagger}\,,\\*
  \sigma \to  L\sigma \mathbb{U}^{\dagger} = \mathbb{U} \sigma R^{\dagger}\,, \qquad&&\qquad
  \sigma^\dagger \to R \sigma^\dagger \mathbb{U}^{\dagger} = \mathbb{U} \sigma^\dagger L^{\dagger}\,, 
  \label{eq:Udef}
\end{eqnarray}
where $L\in SU(4n)_L$, $R\in SU(4n)_R$, and $\mathbb{U}$ is a function of  $L$ and $R$ and the pion fields.
In the construction of invariant Lagrangian terms 
it is convenient to define objects involving the $\sigma$ field that transform 
only with $\mathbb{U}$ and $\mathbb{U}^\dagger$. The two 
possibilities with a single derivative are
\begin{eqnarray}
  \mathbb{V}_{\mu} & = & \frac{i}{2} \left[ \sigma^{\dagger} \partial_\mu
   \sigma + \sigma \partial_\mu \sigma^{\dagger}   \right] \ ,\\
  \mathbb{A}_{\mu} & = & \frac{i}{2} \left[ \sigma^{\dagger} \partial_\mu
   \sigma - \sigma \partial_\mu \sigma^{\dagger}   \right] \ .
\end{eqnarray}

The field that destroys a heavy-light meson can be written as
\begin{equation} \label{eq:H_definition}
  H_{\alpha a} = \frac{1 + \vslash}{2}\left[ \gamma^\mu B^{*}_{\mu \alpha a}
    + i \gamma_5 B_{\alpha a}\right]\ ,
\end{equation}
where $v$ is the meson's velocity, 
$a$ is the combined flavor-taste index of the light quark, and
$\alpha$ is the heavy-quark taste index. 
To avoid confusion with the covariant derivative $\rightvec D_\mu$ introduced below, 
we will use $B$ for now to denote a generic 
pseudoscalar heavy-light meson and $B^*$ to denote the corresponding
vector meson (with $v^\mu B^*_{\mu\alpha a}=0$), even 
though the focus of current all-staggered simulations
is primarily on the $D$ meson system rather than $B$ meson system. The formalism 
developed in this paper applies to both, although $1/m_Q$ corrections are of course larger for $D$'s.
The conjugate field that creates a heavy-light meson is
\begin{equation} \label{eq:Hbar_definition}
  \overline{H}_{a\alpha} \equiv \gamma_0 H^{\dagger}_{a\alpha}\gamma_0 =
  \left[ \gamma^\mu B^{\dagger *}_{\mu a\alpha}
    + i \gamma_5 B^{\dagger}_{a\alpha}\right]\frac{1 + \vslash}{2}\ .
\end{equation}

Under the $SU(2)$ heavy-quark spin symmetry, the heavy-light field
transforms as
\begin{eqnarray}
  H &\to & S H\ , \nonumber\\
  \overline{H} &\to & \overline{H}S^{\dagger}\ ,\label{eq:Hspin}
\end{eqnarray}
with $S\in SU(2)$  acting on Dirac index of the heavy-light field. 
Transformations under the chiral $SU(4n)_L\times SU(4n)_R$ symmetry 
of the light quarks take the form
\begin{eqnarray}
  H &\to & H \mathbb{U}^{\dagger}\ ,\nonumber\\
  \overline{H} &\to & \mathbb{U}\overline{H}\ \eqn{H-transform},
\end{eqnarray}
with $\mathbb{U}\in SU(4n)$ acting on the combined flavor-taste index $a$ in 
\eqs{H_definition}{Hbar_definition}.  Heavy quarks of course do not have
a chiral symmetry, but they do have a vector $SU(4)$ taste symmetry (exact in
the continuum limit), under which
\begin{eqnarray}
  H &\to & V H\ , \nonumber\\
  \overline{H} &\to & \overline{H}V^{\dagger}\ ,\label{eq:Htaste}
\end{eqnarray}
with $V\in SU(4)$ acting on the heavy-quark taste index.

We introduce a  (chirally) covariant derivative that acts on the heavy-light field or its conjugate as
\begin{eqnarray}\label{eq:Ddef}
(H \leftvec D_\mu)_{\alpha b}  = H_{\alpha c} (\leftvec D_\mu)_{cb}  &\equiv& \partial_\mu H_{\alpha b} +
			       i H_{\alpha c}(\mathbb{V}_{\mu})_{cb}\ , \nonumber \\
(\rightvec D_\mu \overline{H})_{b\alpha}  = (\rightvec D_\mu)_{bc} \overline{H}_{c\alpha} &\equiv& \partial_\mu \overline{H}_{b\alpha} - 
					  i (\mathbb{V}_{\mu})_{bc} \overline {H}_{c\alpha}\ ,
\end{eqnarray}
with implicit sums over repeated indices. 

So far $H$ is treated as a $4 \times 4n$ matrix in the taste and the flavor space of quarks. 
Instead of attaching separate indices for the tastes of the light and heavy quarks of the meson,
one can use a single index for the combined meson taste. The field $H$ is then treated
as an $n$-component vector in the flavor
space of the light quark, while each element ($H_i$, $i=1,\dots,n$) 
is a $4\times 4$ matrix in the taste space of the meson, and written as a linear combination
of the 16 taste generators $T_\Xi$, \eq{T_Xi}.
We use Latin indices in the middle of the alphabet $(i, j, ...)$ as pure flavor indices,
and capital Greek letters such as $\Xi$ to indicate meson tastes.  
For example, the $i$th element of the 
field destroying a heavy-light meson in the light flavor space can be represented by 
$H_i = \sum_{\Xi=1}^{16} \frac{1}{2}T_\Xi \,H_{i \Xi } $ and
its conjugate by $\overline{H}_i = \sum_{\Xi=1}^{16} \frac{1}{2}  T_\Xi\, \overline{H}_{i\Xi}$,
where the factors of $\frac{1}{2}$ are inserted to ensure that the fields $H_{i\Xi}$ and 
$\overline{H}_{i\Xi}$ are conventionally normalized.

We can now write down $\cL_1$. As discussed in \secref{power}, lattice corrections
are higher order in the heavy-light system, so at LO we just
have the continuum-like Lagrangian \cite{MAN_WISE,Aubin:StagHL2006} 
\begin{equation}\label{eq:L1}
  \cL_1  =  -i \Tr(\overline{H} H v\negcdot \leftvec D )
  + g_\pi \Tr(\overline{H}H\gamma^{\mu}\gamma_5 
  \mathbb{A}_{\mu}) \ ,
\end{equation}
$\Tr$ means the complete trace over flavor, taste,
and Dirac indices. 
The only difference of $\cL_1$ from the continuum LO Lagrangian is addition of the (implicit)
taste degrees of freedom of light and heavy quarks.
The product $\overline{H}H$ can be treated either as a $4n\times 4n$ matrix in 
the flavor-taste space of the light quarks: 
$(\overline{H}H)_{ab} \equiv \overline{H}_{a \alpha} H_{\alpha b}$ (with an implicit sum over $\alpha$), or equivalently 
as a $n\times n$ matrix in the flavor space of the light quarks,
where each element is itself a $4\times 4$ matrix in the taste space of the meson: 
$(\overline{H}H)_{ij} \equiv  \frac{1}{4}\sum_{\Xi=1}^{16}  \sum_{\Xi'=1}^{16} 
\overline{H}_{i \Xi} H_{j \Xi'} T_\Xi T_{\Xi'}$.
Depending on the situation, one of the notations may be more convenient; 
we must however be careful to be consistent 
in the treatment of  other objects in the same term in the Lagrangian.

For the calculation of the heavy-light decay constants in \secref{fD}, 
the chiral representative of the axial heavy-light current is needed. 
Alternatively, one can work with the left-handed current, whose matrix element between
a pseudoscalar meson and the vacuum is proportional to that of the axial current.
For the current, it is simplest to treat the heavy-light field as a light-flavor vector
whose elements are meson taste matrices.
The left-handed current that destroys a heavy-light meson
of taste $\Xi$ and light flavor $i$ is $j^{\mu,i \Xi}$, which at LO takes the form
\begin{equation}\label{eq:LOcurrent}
  j^{\mu,i \Xi}_{\rm LO} = \frac{\kappa}{2}\; 
  \trDt\bigl(\half T_\Xi\gamma^\mu\left(1-\gamma_5\right)H\sigma^\dagger\lambda^{(i)} \bigr)
\end{equation}
where $\kappa$ is a low-energy constant, and $\lambda^{(i)}$ is a constant row vector that fixes the flavor of the light quark:
$(\lambda^{(i)})_j = \delta_{i j}$. 
This expression for the current is a trivial generalization
of that in Ref.~\cite{MAN_WISE} to include the taste degrees of freedom. 
It can be checked using the spurion analysis introduced in \secref{NLOa2type2} to find the current at
next order.
The decay constant $f_{B_{i \Xi}}$ is defined by the matrix
element
\begin{equation}\label{eq:matrix_element}
  \left\langle 0 \left| j^{\mu,i' \Xi'}
  \right| B_{i \Xi}(v) \right\rangle =i
  f_{B_{i \Xi}}m_{B_{i \Xi}} v^{\mu} \delta_{\Xi \Xi'}\delta_{ii'}\ ,
\end{equation}
where relativistic normalization of the state $|B_{i \Xi}(v) \rangle $ is assumed.
At LO in the heavy-light chiral theory,
$j^{\mu,i' \Xi'}_{\rm LO} = i\kappa v^\mu {B_{i' \Xi'}}$, which gives 
$f_{B_{i \Xi}}^{\rm LO} = \kappa/\sqrt{m_{B_{i \Xi}}}$. Recall that the
factor $\sqrt{m_{B_{i \Xi}}}$ arises from the differences in
normalizations between relativistic and non-relativistic states.

\subsection{Next-to-leading-order terms in the continuum}
\label{sec:NLO-CONTINUUM}

In the continuum, the NLO terms are of two types: those formed by only adding derivatives
to LO terms ($\cL_{2,k}$, $\cL_{3,k}$, 
$j^{\mu,i \Xi}_1 $,  and $j^{\mu,i \Xi}_{2,k}$), and those that involve a mass spurion
($\cL_{2,m}$, $\cL_{3,m}$, and $j^{\mu,i \Xi}_{2,m}$).  The former
are not to our knowledge cataloged completely in the literature, and in any case are 
irrelevant to the heavy meson mass and decay constant to the order we are working:  
Additional derivatives acting on a heavy-light field vanish
on shell ($k=0$), while those on
the light fields contribute only to tree-level diagrams with external pions.
We therefore follow \rcite{Aubin:StagHL2006}, and simply list some representative terms in
$\cL_{2,k}$, $\cL_{3,k}$,
$j^{\mu,i \Xi}_1 $,  $j^{\mu,i \Xi}_{2,k}$.
We have
\begin{eqnarray} 
\cL_{2,k} &=& \frac{i\epsilon_1}{\Lambda_\chi}\Tr\left((v\cdot \rightvec D\, \overline{H}H 
	  - \overline{H}H v\cdot\leftvec D)\,\gamma_\mu \gamma_5 \mathbb{A}^\mu\right)
	  + \frac{\epsilon_2}{\Lambda_\chi}\Tr\left( \overline{H} H (v\cdot \leftvec D\,)^2 \right)   + \dots 
	  \label{eq:L2k} \\
\cL_{3,k} &=& 
	  \frac{\epsilon_3}{\Lambda_\chi^2}\Tr\left(\overline{H}H \gamma_\mu \gamma_5 
	  (v\cdot \rightvec D\,)^2 \mathbb{A}^\mu\right) 
	  + \frac{\epsilon_4}{\Lambda_\chi^2}\Tr\left(\overline{H} H \rightvec \Dslash\,\gamma_5    \, v\cdot \rightvec D \, 
	  v\cdot \mathbb{A}\right) + \dots 
	  \label{eq:L3k} \\
j^{\mu,i \Xi}_{1,k} &=& \frac{i\kappa_1}{\Lambda_\chi}\; 
		      \trDt\bigl(\half T_\Xi\gamma^\mu \left(1\!-\!\gamma_5\right) H v\cdot \leftvec D\,
		    \sigma^\dagger \lambda^{(i)} \bigr)  \nonumber \\
		      &&\quad + \frac{\kappa_2}{\Lambda_\chi}\; 
		      \trDt\bigl(\half T_\Xi\gamma^\mu \left(1\!-\!\gamma_5\right) H\, 
		    v\cdot  \mathbb{A}\, \sigma^\dagger \lambda^{(i)} \bigr) +\dots 
		    \label{eq:j1k} \\		
j^{\mu,i \Xi}_{2,k} &=& \frac{\kappa_3}{\Lambda^2_\chi}\; 
		      \trDt\bigl(\half T_\Xi\gamma^\mu \left(1\!-\!\gamma_5\right) H (v\cdot \leftvec D\,)^2
		    \sigma^\dagger \lambda^{(i)} \bigr) \nonumber  \\
		      &&\quad + \frac{i\kappa_4}{\Lambda^2_\chi}\; 
		      \trDt\bigl(\half T_\Xi\gamma^\mu \left(1\!-\!\gamma_5\right) H\, 
		    v\cdot \rightvec D\, v\cdot  \mathbb{A}\, \sigma^\dagger \lambda^{(i)} \bigr) +\dots \label{eq:j2k}
\end{eqnarray}
where the constants $\epsilon_i, \kappa_j$ are taken to be real and dimensionless, 
$\Lambda_\chi$ is the chiral scale, and
\begin{equation}
\rightvec D_\nu \mathbb{A}_\mu \equiv \partial_\nu \mathbb{A}_\mu   -i [\mathbb{V}_\nu, \mathbb{A}_\mu] \ .
\end{equation}
The only difference from \rcite{Aubin:StagHL2006} is 
a small change of notation because of the taste degree of freedom of the heavy quark:
Here the current has meson taste $\Xi$ and light flavor fixed by $\lambda^{(i)}$; whereas in 
\rcite{Aubin:StagHL2006} the current had only light-quark taste and flavor, both of which
were fixed by $\lambda^{(i)}$.

The terms induced by single insertions of the light quark mass spurions also follow
directly from \rcite{Aubin:StagHL2006}.  They are:
\begin{eqnarray}
\cL_{2,m} & = &  2\lambda_1 \Tr\left(\overline{H} H\cM^+\right) 
	    + 2\lambda'_1 \Tr\left(\overline{H} H\right)\Tr\left(\cM^+\right)   \ , \label{eq:L2m} \\
\cL_{3,m} & = & 
	  i k_1 \Tr\left( \overline{H}H  v\negcdot \leftvec D\, \cM^+ - v\negcdot \rightvec D \,
	  \overline{H}H\, \cM^+ \right) \nonumber \\
	  &&+ ik_2 \Tr\left( \overline{H} H  v\negcdot \leftvec D -v\negcdot \rightvec D \,
	  \overline{H}H\right) \Tr(\cM^+) \nonumber\\  &&{}+ 
	    k_3 \Tr\left( \overline{H} H  \gamma_\mu\gamma_5\{\mathbb{A}^\mu ,
	    \cM^+\}\right) +  k_4 \Tr\left( \overline{H}
	      H \gamma_\mu\gamma_5 \mathbb{A}^\mu \right) \Tr(\cM^+) \nonumber\\
	    &&{} + k_5 \Tr\left( \overline{H} H  \gamma_\mu\gamma_5\right)\Tr\left(\mathbb{A}^\mu \cM^+\right) +
	    k_6 \Tr\left( \overline{H} H \gamma_\mu [\mathbb{A}^\mu , \cM^-] \right) \ , \label{eq:L3m} \\
j^{\mu,i \Xi}_{2,m} & = & 
		      \rho_1\, \trDt\left(\half T_\Xi\gamma^\mu (1-\gamma_5) H   \cM^+ \sigma^\dagger  \lambda^{(i)} \right)
		      + \rho_2\, \trDt\left(\half T_\Xi\gamma^\mu (1-\gamma_5) H \sigma^\dagger \lambda^{(i)} \right) \Tr(\cM^+) \nonumber \\
		    &&\hspace{-9mm}+ \rho_3\, \trDt\left(\half T_\Xi\gamma^\mu (1-\gamma_5) H\cM^- \sigma^\dagger \lambda^{(i)} \right)  
		    + \rho_4\, \trDt\left(\half T_\Xi\gamma^\mu (1-\gamma_5) H\sigma^\dagger \lambda^{(i)} \right) \Tr(\cM^-), \eqn{j2m}
\end{eqnarray}
where $\cM^\pm  =  \half \left(\sigma \cM\sigma
  \pm \sigma^{\dagger} \cM\sigma^{\dagger}\right)$ are the light-quark mass spurions.

\section{Taste Symmetry Breaking}\label{sec:4quark-ops}

Taste violations first appear at $\cO(a^2)$. In the SET, they are described by four-quark
(dimension six) operators, which are generated by gluon exchange with total momenta $\sim\! \pi/a$
between two quark lines.  The gluons can change the taste, spin, and color of the quark
line, but not its flavor, so the operators take the form of products of two
quark bilinears, where each bilinear is made of quark and antiquark fields of
a single flavor.  In the current case, there are three generic classes of
four-quark operators: where both bilinears are of light quarks, 
where one bilinear is light and the other heavy,
and where both bilinears heavy.  We write
\begin{eqnarray}\label{eq:SETa_ll}
a^2\cO^{ll}_{ss'tt'}&  = & c_1  a^2\;\overline{q}_l(\gamma_s\otimes\xi_t)q_l 
                            \; \overline{q}_{l'}(\gamma_{s'}\otimes\xi_{t'})q_{l'} \ , \\ 
a^2\cO^{lh}_{ss'tt'} &= & c_2  a^2\;\overline{q}_l(\gamma_s\otimes\xi_t)q_l 
                              \;   \overline{q}_h(\gamma_{s'}\otimes\xi_{t'})q_h  \ ,
\eqn{SETa_lh} \\ 
a^2\cO^{hh}_{ss'tt'} &= & c_3  a^2\;\overline{q}_h(\gamma_s\otimes\xi_t)q_h 
                                \; \overline{q}_{h'}(\gamma_{s'}\otimes\xi_{t'})q_{h'}
\eqn{SETa_hh} \ .
\end{eqnarray}
where $l$ and $h$ refer to light and heavy quarks, respectively;
$s,s'$ label spins; and $t,t'$ label tastes. The light quark labels $l$ and $l'$ are
summed over; only a single heavy quark flavor is considered.
Color indices, which may be contracted
in different ways, are omitted 
because they have no effect on the 
chiral operators generated. 
The operators
in \eqsthru{SETa_ll}{SETa_hh} are schematic; they stand for the whole
set of possible four-quark operators with the given flavor structure. 
Similarly, 
each coefficient $c_i$ represents a set of coefficients of the operators.

The staggered symmetries impose the following constraints on the possible 
operators\footnote{See \rcite{Bazavov:2009bb} for a pedagogical review; we follow it closely.}
\begin{eqnarray}\label{eq:U1eps}
U(1)_\epsilon \textrm{ symmetry} & \Rightarrow & 
\{\gamma_5\otimes\xi_5, \gamma_s\otimes\xi_t \} = 0 \ , \\
\textrm{shift symmetry} & \Rightarrow & \xi_t = \xi_{t'} \ , \eqn{shift} \\
\textrm{rotational and parity symmetries} & \Rightarrow & \gamma_t = \gamma_{t'} 
\eqn{rot-and-parity}\ . 
\end{eqnarray}

At this point the lattice spacing $a$ has simply become a parameter in the
continuum SET theory.  We can therefore use the fact  that the heavy quark
mass $m_Q$ is large compared to $\Lambda_{QCD}$ to replace the field $q_h$ in
\eqs{SETa_lh}{SETa_hh} with the HQET field $Q$, \eq{Qdef}. Making in addition the simplifications
implied by \eqs{shift}{rot-and-parity}, we have
\begin{eqnarray}\label{eq:SET_ll}
a^2\cO^{ll}_{st}&  = & c_1  a^2\;\overline{q}_l(\gamma_s\otimes\xi_t)q_l 
                            \; \overline{q}_{l'}(\gamma_{s}\otimes\xi_{t})q_{l'} \ , \\ 
a^2\cO^{lh}_{st} &= & c_2  a^2\;\overline{q}_l(\gamma_s\otimes\xi_t)q_l 
                              \;   \overline{Q}(\gamma_{s}\otimes\xi_{t})Q  \ ,
\eqn{SET_lh} \\ 
a^2\cO^{hh}_{st} &= & c_3  a^2\;\overline{Q}(\gamma_s\otimes\xi_t)Q 
                                \; \overline{Q}(\gamma_{s}\otimes\xi_{t})Q
\eqn{SET_hh} \ .
\end{eqnarray}

The operators can be further separated into type A and type B operators \cite{LEE_SHARPE}, 
which are distinguished by whether they break continuum  Euclidean rotation symmetry.
This breaking occurs when there are 
indices that are common to both the spin and taste matrices, thereby coupling spin and taste. 
Type-A operators 
are invariant under rotation symmetry, while 
type-B operators break it.  Both types of operators 
break $SU(4)$ taste symmetry. Type-A operators are, however, 
invariant under an $SO(4)$ 
taste subgroup, as well as the $SO(4)$ of space-time rotations, whereas type-B operators
are invariant only under combined $90^\circ$ rotations of both spin and taste.
There are a total of twelve type-A operators that are named
by the spin $\otimes$ taste of their bilinears \cite{LEE_SHARPE}:
\begin{eqnarray}
 && [S\times A],\  [S\times V],\  [A\times S],\  [V\times S],\  [P\times A],\  [P\times V],\ 
 \nonumber \\
             && [A\times P],\  [V\times P],\  [T\times V],\  [T\times A],\  [V\times T],\  [A\times T]\ .
\label{eq:SFFA}
\end{eqnarray}  
Each operator will also have the superscript $ll$, $lh$, or $hh$ to denote 
its flavor.
Thus, for example
\begin{equation}\eqn{example-A}
 [T\times A]^{lh} \equiv a^2\, \overline{q}_l(\gamma_{\mu\nu}\otimes \xi_{\lambda 5})q_l 
                                \; \overline{Q}(\gamma^{\nu\mu}\otimes \xi_{5\lambda})Q  \ ,
\end{equation}
where $\gamma_{\mu\nu} \equiv (1/2)[\gamma_\mu,\gamma_\nu]$, and we use 
Minkowski  gamma
matrices for convenience, corresponding to the fact that we have chosen to
write the chiral Lagrangian ultimately in Minkowski space. Taste matrices
remain Euclidean, as in \eq{T_Xi}.  Summation
over the twice-repeated indices $\mu,\nu,\lambda$ is implied.

There are four type-B operators:
\begin{equation}
 [T_\mu\times V_\mu],\  [T_\mu\times A_\mu],\  [V_\mu\times T_\mu],\  [A_\mu\times T_\mu],
 \label{eq:SFFB}
\end{equation}
where $\mu$ is the common index that appears four times.  For example, we have
\begin{equation}\eqn{example-B}
 [A_\mu\times T_\mu]^{ll} \equiv a^2\, \sum_{\mu} \overline{q}_l(i\gamma_{\mu}\gamma_5\otimes i\xi_{\mu \nu})q_l 
                                \; \overline{q}_{l'}(i\gamma^{\mu}\gamma_5\otimes i\xi_{\mu \nu})q_{l'}  \ .
\end{equation}
The index $\nu$, which appears twice, obeys the summation convention, while the sum over 
an index like $\mu$, which appears four times, is shown explicitly here and below.

We now consider the chiral operators that correspond to the SET/HQET
operators, \eqs{SET_ll}{SET_hh}. The light-light operators, \eq{SET_ll}, are 
(trivially) invariant under the heavy-quark taste symmetry, while breaking
the light-quark taste symmetry, leading to the 
NLO terms in the Lagrangian and current denoted by
$\cL_{2,a^2}^{A1}$, $\cL_{2,a^2}^{B1}$, $\cL_{3,a^2}^{A1}$, $\cL_{3,a^2}^{B1}$, $j^{\mu,i \Xi}_{2,a^2,A1}$ 
and $j^{\mu,i \Xi}_{2,a^2,B1}$ in
\eqsthru{L2}{current2}.  They are summarized in the following subsection. 
The light-heavy operators, \eq{SET_lh}, break both the light-quark and
heavy-quark taste symmetries. 
These operators lead to the terms  
denoted by
$\cL_{2,a^2}^{A2}$, $\cL_{2,a^2}^{B2}$, $\cL_{3,a^2}^{A2}$, $\cL_{3,a^2}^{B2}$, $j^{\mu,i \Xi}_{2,a^2,A2}$ 
and $j^{\mu,i \Xi}_{2,a^2,B2}$ in \eqsthru{L2}{current2}, and are discussed in \secref{NLOa2type2}.
Although the heavy-heavy operators, \eq{SET_hh}, break the heavy taste symmetry, they 
do not result in any new chiral operators in the heavy-light chiral Lagrangian or
current, for reasons we discuss at the end of \secref{NLOa2type2}.

\subsection{Discretization errors at NLO:  Light-taste breaking terms}
\label{sec:NLOa2type1}

The
light-light operators in \eq{SET_ll} are trivially invariant under the
heavy-quark spin symmetry, in addition to the heavy-quark taste symmetry. 
Either symmetry alone is enough to
guarantee that all corresponding Lagrangian operators are composed of the product
$\overline{H} H$. This means that operators determined in 
\rcite{Aubin:StagHL2006} from the light-light four-quark operators
can be taken over without change even
though the heavy quarks considered there had no taste degree of freedom. 
Similarly, 
the spin symmetry alone requires that the left-handed current is constructed
from the combination $\gamma^\mu(1-\gamma_5)H$, and the heavy-quark taste symmetry
provides no fundamentally new information. Thus the current can also be taken over
from \rcite{Aubin:StagHL2006}, although in this case one needs
the same minor notational change to accommodate the heavy-quark taste degree of freedom that
we have used above in \eqsthree{j1k}{j2k}{j2m}.
We have also found it necessary to change a few symbols from those used in 
\rcite{Aubin:StagHL2006} in order to avoid conflict with notation in the present paper.
Moreover, we have discovered a few new terms 
that were missed in that 
reference, and have dropped a few terms  that 
are not independent or are absent for other reasons.  The changes have no effect on
existing calculations in \rhmschpt: the
heavy-light leptonic decay constant \cite{Aubin:StagHL2006} and the semileptonic form factors
for heavy-light meson decays to light 
\cite{Aubin:StagHL2007} or heavy-light \cite{Laiho:2005ue} mesons.

For type-A operators, the contributions to the chiral Lagrangian are
\begin{equation}
\cL^{A1}_{2,a^2}  = a^2\sum_{k=1}^8 \Biggl\{K^{A1}_{1,k} \Tr\left(\overline{H} H
		  \cO^{A1,+}_k\right)
		  +  K^{A1}_{2,k}\Tr\left(\overline{H} H\right)
			\Tr(  \cO^{A1,+}_k) \Biggr\} \,  \label{eq:L2A1}
\end{equation}
and
\begin{eqnarray}
\cL^{A1}_{3,a^2} & = &   a^2\sum_{k=1}^8\Biggl\{
	ic^{A1}_{1,k} \Tr\left( \overline{H}H  v\negcdot \leftvec D\, \cO^{A1,+}_k - v\negcdot \rightvec D \,
	\overline{H}H\, \cO^{A1,+}_k  \right) \nonumber \\
      &&\hspace{1.2truecm}+ i c^{A1}_{2,k} 
	\Tr\left( \overline{H} H  v\negcdot \leftvec D -v\negcdot \rightvec D \,
	 \overline{H}H\right)  \Tr(\cO^{A1,+}_k) \nonumber\\
      &&\hspace{1.2truecm}+ c^{A1}_{3,k} \Tr\left( \overline{H}
	H \gamma_\mu\gamma_5\{\mathbb{A}^\mu, 	\cO^{A1,+}_k\}\right)
	+ c^{A1}_{4,k} \Tr\left( \overline{H} H \gamma_\mu\gamma_5\mathbb{A}^\mu   \right)
	\Tr(\cO^{A1,+}_k) \nonumber\\
      &&\hspace{1.2truecm}+ c^{A1}_{5,k}\Tr\left( \overline{H}
	H \gamma_\mu\gamma_5\right)\Tr(\mathbb{A}^\mu  \cO^{A1,+}_k)+ 
        c^{A1}_{6,k} \Tr\left( \overline{H} H \gamma_\mu[\mathbb{A}^\mu,
	\cO^{A1,-}_k]\right) \nonumber \\
      &&\hspace{1.2truecm}+ c^{A1}_{7,k} \Big( \Tr\big(\overline{H}
	H \gamma_\mu\gamma_5 P^{A1}_k \mathbb{A}^\mu 
	{\tilde P^{A1}_k}\big) + {\rm p.c.} \Big)  \nonumber\\
      &&\hspace{1.2truecm}+ c^{A1}_{8,k} \Big( \Tr\big(\overline{H}
	H \gamma_\mu\gamma_5 P^{A1}_k\big)\Tr\big( \mathbb{A}^\mu
	{\tilde P^{A1}_k}\big) + {\rm p.c.} \Big) \Biggr\}  \nonumber\\
      &&+a^2\sum_{k=2,5,7,8}c^{A1}_{9,k} \Big( \Tr\big(\overline{H}
	H \gamma_\mu P^{A1}_k\mathbb{A}^\mu {\tilde P^{A1}_k}\big) +{\rm p.c.}\Big)\nonumber \\
      &&+a^2\sum_{k=1,2,6,7}c^{A1}_{10,k} \Big( \Tr\big(\overline{H}
	H \gamma_\mu P^{A1}_k\big)\Tr\big( \mathbb{A}^\mu {\tilde P^{A1}_k}\big)+{\rm p.c.}\Big) \ .  \label{eq:L3A1}
\end{eqnarray}
where ${\rm p.c.}$ denotes the parity conjugate; for example, $\sigma_{{\rm p.c.}}=\sigma^\dagger$.  
Taste violations are encoded in the operators
\begin{eqnarray}\label{eq:OA1}
  \cO^{A1,\pm}_1 & = &  (\sigma\xi^{(n)}_5\Sigma^{\dagger}\xi^{(n)}_5\sigma 
  \pm {\rm p.c.})\nonumber \\*
  \cO^{A1,\pm}_2 & = & 
  \left[ (\sigma\xi^{(n)}_{\nu}\sigma)\Tr(\xi^{(n)}_\nu\Sigma)
    \pm {\rm p.c.}\right]\nonumber \\*
  \cO^{A1,\pm}_3 & = & ( \sigma\xi^{(n)}_{\nu}\Sigma
    \xi^{(n)}_\nu\sigma \pm {\rm p.c.}) \nonumber \\*
  \cO^{A1,\pm}_4 & = & ( \sigma\xi^{(n)}_{\nu 5}\Sigma
    \xi^{(n)}_{5\nu}\sigma \pm {\rm p.c.}) \nonumber \\*
  \cO^{A1,\pm}_5 & = & 
  \left[ (\sigma\xi^{(n)}_{\nu}\sigma)\Tr(\xi^{(n)}_\nu\Sigma^\dagger)
  \pm {\rm p.c.} \right]\nonumber \\*
  \cO^{A1,\pm}_6 & = &  
  (\sigma\xi^{(n)}_{\mu\nu}\Sigma^{\dagger}
  \xi^{(n)}_{\nu\mu}\sigma \pm {\rm p.c.})  \nonumber \\*
  \cO^{A1,\pm}_7 & = & 
  \left[ (\sigma \xi^{(n)}_{\nu5}\sigma)\Tr( \xi^{(n)}_{5\nu}\Sigma)
    \pm {\rm p.c.}\right]\nonumber\\*
  \cO^{A1,\pm}_8 & = & 
  \left[ (\sigma \xi^{(n)}_{\nu5}\sigma)\Tr(\xi^{(n)}_{5\nu}\Sigma^\dagger)
   \pm {\rm p.c.} \right]\ ,
\end{eqnarray}
and
\begin{eqnarray}\label{eq:PA}
  & P^{A1}_1  =   \sigma\xi^{(n)}_5\sigma^\dagger\;,&\quad {\tilde P^{A1}_1} \equiv 
  (P^{A1}_1)_{{\rm p.c.}}=\sigma^\dagger\xi^{(n)}_5\sigma \nonumber \\*
  & P^{A1}_2  =   \sigma\xi^{(n)}_5\sigma^\dagger\;,&\quad {\tilde P^{A1}_2} \equiv 
  P^{A1}_2 \nonumber \\*
  & P^{A1}_3  =   \sigma\xi^{(n)}_\nu\sigma\;,&\quad {\tilde P^{A1}_3} \equiv 
  P^{A1}_3 \nonumber \\*
  & P^{A1}_4  =   i\sigma\xi^{(n)}_{\nu5}\sigma\;,&\quad {\tilde P^{A1}_4} \equiv 
  P^{A1}_4 \nonumber \\*
  & P^{A1}_5  =   \sigma\xi^{(n)}_\nu\sigma\;,&\quad {\tilde P^{A1}_5} \equiv 
  (P^{A1}_5)_{{\rm p.c.}}   =   \sigma^\dagger\xi^{(n)}_\nu\sigma^\dagger \nonumber \\*
  & P^{A1}_6  =   i\sigma\xi^{(n)}_{\lambda\nu}\sigma^\dagger\;,&\quad {\tilde P^{A1}_6} \equiv 
  (P^{A1}_6)_{{\rm p.c.}}=-i\sigma^\dagger\xi^{(n)}_{\nu\lambda}\sigma \nonumber \\*
  & P^{A1}_7  =   i\sigma\xi^{(n)}_{\lambda\nu}\sigma^\dagger\;,&\quad {\tilde P^{A1}_7} \equiv 
  P^{A1}_7 \nonumber \\*
  & P^{A1}_8  =   i\sigma\xi^{(n)}_{\nu5}\sigma\;,&\quad {\tilde P^{A1}_8} \equiv 
  (P^{A1}_8)_{{\rm p.c.}}=-i\sigma^\dagger\xi^{(n)}_{5\nu}\sigma^\dagger\;. 
\end{eqnarray}

For the current, we have
\begin{eqnarray}
j^{\mu,i \Xi}_{2,a^2,A1} & = & a^2\sum_{k=1}^8
      \Biggl\{ r^{A1}_{1,k}\, \trDt\left(\half T_\Xi\gamma^\mu (1-\gamma_5) H \cO^{A1,+}_k \sigma^\dagger \lambda^{(i)} \right )\nonumber \\
      &&\qquad + r^{A1}_{2,k}\, \trDt\left(\half T_\Xi\gamma^\mu (1-\gamma_5) H \sigma^\dagger \lambda^{(i)} \right) \Tr(\cO^{A1,+}_k)
       + r^{A1}_{3,k}\, \trDt\left(\half T_\Xi\gamma^\mu (1-\gamma_5) H \cO^{A1,-}_k \sigma^\dagger \lambda^{(i)} \right ) \nonumber \\
       &&\qquad + r^{A1}_{4,k}\, \trDt\left(\half T_\Xi\gamma^\mu (1-\gamma_5) H \sigma^\dagger \lambda^{(i)} \right)
       \Tr(\cO^{A1,-}_k) \Biggr\}\ . \label{eq:j2A1}
\end{eqnarray}

Similarly, for type-B operators, we have: 
\begin{eqnarray}
\cL^{B1}_{2,a^2}   &=&  a^2\sum_\mu\sum_{k=1}^3
	      \Biggl\{ K^{B1}_{1,k}v_\mu v^\mu \Tr(\overline{H} H
	      \cO^{B1,+}_{\mu,k}) + K^{B1}_{2,k} v_\mu v^\mu  \Tr(\overline{H} H)
	      \Tr(\cO^{B1,+}_{\mu,k})  \nonumber \\
	      &&\qquad + K^{B1}_{3,k} v_\mu  \Tr(\overline{H} H \gamma^\mu \gamma_5
	      \cO^{B1,-}_{\mu,k}) + K^{B1}_{4,k} v_\mu  \Tr(\overline{H} H \gamma^\mu\gamma_5)
	      \Tr(\cO^{B1,-}_{\mu,k}) \Biggr\}\ ,\label{eq:L2B1}  
\end{eqnarray}
where
\begin{eqnarray}\label{eq:OB1}
    \cO^{B1,\pm}_{\mu, 1} & = & 
    (\sigma\xi^{(n)}_{\mu\lambda}\Sigma^\dagger\xi^{(n)}_{\lambda\mu}\sigma) 
    \pm {\rm p.c.}\ ,\nonumber \\ 
    \cO^{B1,\pm}_{\mu, 2} & = &  
    (\sigma\xi^{(n)}_\mu\sigma)
    \Tr(\xi^{(n)}_\mu\Sigma^\dagger) \pm {\rm p.c.} \ ,\nonumber \\
    \cO^{B1,\pm}_{\mu, 3} & = &  
    (\sigma\xi^{(n)}_{\mu5}\sigma)
    \Tr(\xi^{(n)}_{5\mu}\Sigma^\dagger) \pm {\rm p.c.} \ .
\end{eqnarray}
Note that the above operators explicitly depend on $\mu$, and there is no summation over this index in their definition. 
(We do sum over $\lambda$.) The sum over $\mu$ is shown explicitly in \eq{L2B1}. 
The terms proportional to $K^{B1}_{3,k}$ and $K^{B1}_{4,k}$
in \eq{L2B1}, which have the form of a product of a parity-odd combination of the 
heavy-light mesons times a parity-odd combination of the light mesons, were
omitted in \rcite{Aubin:StagHL2006}. They are unlikely to be important in practical 
calculations in either \rhmschpt\ and \aschpt\ since their 
first contribution is a NLO correction to the $B$-$B^*$-$\pi$ vertex.

There are many terms in $\cL^{B1}_{3,a^2}$, so we separate it for
convenience into two parts:
\begin{equation}
\cL^{B1}_{3,a^2}   = \cL^{B1,O}_{3,a^2}  + \cL^{B1,P}_{3,a^2} \ . \label{eq:L3B1} 
\end{equation}
We then have
\begin{eqnarray}
\cL^{B1,O}_{3,a^2} &=&  a^2\sum_\mu\sum_{k=1}^3
	      \Biggl\{ ic^{B1}_{1,k} \Tr\left( \overline{H}H  v^\mu \leftvec D_\mu\, \cO^{B1,+}_{\mu,k} - v^\mu \rightvec D_\mu \,
	      \overline{H}H\, \cO^{B1,+}_{\mu,k} \right) \nonumber \\
	&&\hspace{0.7truecm}+ i c^{B1}_{2,k} \Tr\left( \overline{H} H  v^\mu \leftvec D_\mu -v^\mu \rightvec D_\mu \,
	      \overline{H}H\right) \Tr(\cO^{B1,+}_{\mu,k}) \nonumber\\
	&&\hspace{0.7truecm}+ c^{B1}_{3,k} \Tr\left( \overline{H}
	      H \gamma_\mu\gamma_5\{\mathbb{A}^\mu, \cO^{B1,+}_{\mu,k}\}\right)
	      + c^{B1}_{4,k} \Tr\left( \overline{H} H \gamma_\mu\gamma_5\mathbb{A}^\mu   \right)
	      \Tr(\cO^{B1,+}_{\mu,k}) \nonumber\\&&\hspace{0.7truecm}+
	      c^{B1}_{5,k}\Tr\left( \overline{H} H \gamma_\mu\gamma_5\right)\Tr(\mathbb{A}^\mu
	      \cO^{B1,+}_{\mu,k})+ c^{B1}_{6,k} \Tr\left( \overline{H}
	      H \gamma_\mu[\mathbb{A}^\mu, \cO^{B1,-}_{\mu,k}]\right)\nonumber \\
	&&\hspace{0.7truecm}+ ic^{B1}_{7,k} \, v_\mu v^\mu 
	      \Tr\left( \overline{H}H  v\negcdot \leftvec D\, \cO^{B1,+}_{\mu,k} - v\negcdot \rightvec D \,
	      \overline{H}H\, \cO^{B1,+}_{\mu,k} \right) \nonumber \\
	&&\hspace{0.7truecm}+ i c^{B1}_{8,k}\, v_\mu v^\mu  
	      \Tr\left( \overline{H} H  v\negcdot \leftvec D -v\negcdot \rightvec D \,
	      \overline{H}H\right) \Tr(\cO^{B1,+}_{\mu,k}) \nonumber\\
	&&\hspace{0.7truecm}+ c^{B1}_{9,k} \, v_\mu v^\mu \Tr\left( \overline{H}
	      H \gamma_\nu\gamma_5\{\mathbb{A}^\nu, \cO^{B1,+}_{\mu,k}\}\right)
	      + c^{B1}_{10,k} \, v_\mu v^\mu \Tr\left( \overline{H} H \gamma_\nu\gamma_5\mathbb{A}^\nu   \right)
	      \Tr(\cO^{B1,+}_{\mu,k})   \nonumber\\&&\hspace{0.7truecm}+
	      c^{B1}_{11,k}\, v_\mu v^\mu \Tr\left( \overline{H}  H \gamma_\nu\gamma_5\right)\Tr(\mathbb{A}^\nu
	      \cO^{B1,+}_{\mu,k})+   c^{B1}_{12,k} \, v_\mu v^\mu \Tr\left( \overline{H}
	      H \gamma_\nu[\mathbb{A}^\nu, \cO^{B1,-}_{\mu,k}]\right)\nonumber \\
	&&\hspace{0.7truecm}+  c^{B1}_{13,k}\, v^\mu \Tr\left( \overline{H}
	      H \gamma_\mu\gamma_5\{v\negcdot \mathbb{A},   \cO^{B1,+}_{\mu,k}\}\right)
	      + c^{B1}_{14,k}\, v^\mu \Tr\left( \overline{H} H \gamma_\mu\gamma_5\, v\negcdot \mathbb{A}   \right)
	      \Tr(\cO^{B1,+}_{\mu,k}) \nonumber\\&&\hspace{0.7truecm}+
	      c^{B1}_{15,k}\, v^\mu \Tr\left( \overline{H} H \gamma_\mu\gamma_5\right)\Tr(v\negcdot \mathbb{A}\,
	      \cO^{B1,+}_{\mu,k}) +  
	      c^{B1}_{19,k} v^\mu \Tr\left( \overline{H} H \gamma_{\mu\nu}\{\mathbb{A}^\nu,
	      \cO^{B1,-}_{\mu,k}\}\right) 
	      \nonumber \\
	&&\hspace{0.7truecm}+ c^{B1}_{20,k} v^\mu \Tr\left( \overline{H}
	      H \gamma_{\mu\nu}\right)\Tr\left(\mathbb{A}^\nu \cO^{B1,-}_{\mu,k}\right) 
        \!\Biggr\},  \label{eq:L3B1O}
\end{eqnarray}
and
\begin{eqnarray}
\cL^{B1,P}_{3,a^2} &=&  a^2\sum_\mu \Biggl\{ \sum_{k=1}^4 \bigg[
	      c^{B1}_{21,k} \Big(\Tr\big( \overline{H} H \gamma_\mu\gamma_5 P^{B1}_{\mu,k} \mathbb{A}^\mu
	      \tilde P^{B1}_{\mu,k}\big) +{\rm p.c.}\Big) \nonumber\\
	&&\hspace{2.3truecm}+ c^{B1}_{22,k} \Big(\Tr\big( \overline{H} H \gamma_\mu\gamma_5 P^{B1}_{\mu,k} \big)
	      \Tr\big(\mathbb{A}^\mu \tilde P^{B1}_{\mu,k}  \big) +{\rm p.c.}\Big)  \nonumber\\
	&&\hspace{2.3truecm}+  c^{B1}_{23,k} v_\mu v^\mu \Big(\Tr\big( \overline{H}
	      H \gamma_\nu\gamma_5 P^{B1}_{\mu,k} \mathbb{A}^\nu  
	      \tilde P^{B1}_{\mu,k}\big) +{\rm p.c.}\Big) \nonumber\\
	&&\hspace{2.3truecm}+  c^{B1}_{24,k} v_\mu v^\mu \Big(\Tr\big( \overline{H}
	      H \gamma_\nu\gamma_5 P^{B1}_{\mu,k} \big) 
	      \Tr\big(\mathbb{A}^\nu \tilde P^{B1}_{\mu,k}  \big) +{\rm p.c.}\Big) \nonumber\\
	&&\hspace{2.3truecm}+   c^{B1}_{25,k} v^\mu \Big(\Tr\big( \overline{H}
	      H \gamma_\mu\gamma_5 P^{B1}_{\mu,k} v\negcdot\mathbb{A}\,
	      \tilde P^{B1}_{\mu,k}\big) +{\rm p.c.}\Big) \nonumber\\
	&&\hspace{2.3truecm}+  c^{B1}_{26,k} v^\mu \Big(\Tr\big( \overline{H}
	      H \gamma_\mu\gamma_5 P^{B1}_{\mu,k} \big) 
	      \Tr\big(v\negcdot\mathbb{A}\, \tilde P^{B1}_{\mu,k}  \big) +{\rm p.c.}\Big) 
	      \bigg] \nonumber\\
	&&\hspace{0.8truecm}+ \sum_{k=2,3,4} \bigg[  c^{B1}_{29,k} \Big(\Tr\big( \overline{H}
	      H \gamma_\mu P^{B1}_{\mu,k} \mathbb{A}^\mu \tilde P^{B1}_{\mu,k}\big) +{\rm p.c.}\Big) \nonumber\\
	&&\hspace{2.3truecm}+ c^{B1}_{30,k} v_\mu v^\mu \Big(\Tr\big( \overline{H}
	      H \gamma_\nu P^{B1}_{\mu,k} \mathbb{A}^\nu
	      \tilde P^{B1}_{\mu,k}\big) +{\rm p.c.}\Big)\bigg] \nonumber\\
	&&\hspace{0.8truecm}+ \sum_{k=1,4} \bigg[ c^{B1}_{31,k} \Big(\Tr\big( \overline{H}
	      H \gamma_\mu P^{B1}_{\mu,k} \big) 
	      \Tr\big(\mathbb{A}^\mu \tilde P^{B1}_{\mu,k}  \big) +{\rm p.c.}\Big) \nonumber\\
	&&\hspace{2.3truecm}+ c^{B1}_{32,k} v_\mu v^\mu \Big(\Tr\big( \overline{H}
	      H \gamma_\nu P^{B1}_{\mu,k} \big) 
	      \Tr\big(\mathbb{A}^\nu \tilde P^{B1}_{\mu,k}  \big) +{\rm p.c.}\Big) \bigg] \nonumber\\
	&&\hspace{0.8truecm}+ c^{B1}_{33,1} v^\mu \Big( \Tr\big( \overline{H}
	      H \gamma_{\mu\nu} P^{B1}_{\mu,1}
	      \mathbb{A}^\nu \tilde P^{B1}_{\mu,1} \big) +{\rm p.c.}\Big) \nonumber\\
	&&\hspace{0.8truecm}+ \sum_{k=2,3} \bigg[
	      c^{B1}_{34,k} v^\mu \Big( \Tr\big( \overline{H}
	      H \gamma_{\mu\nu} P^{B1}_{\mu,k}\big) 
	      \Tr\big( \mathbb{A}^\nu \tilde P^{B1}_{\mu,k} \big) +{\rm p.c.}\Big)\bigg] 
	\Biggr\} \ , \label{eq:L3B1P}
\end{eqnarray}
where
\begin{eqnarray}\label{eq:PB1}
  & P^{B1}_{\mu,1}  =   i\sigma\xi^{(n)}_{\mu\lambda}\sigma^\dagger\;,&\quad {\tilde P^{B1}_{\mu,1}} \equiv 
    (P^{B1}_{\mu,1})_{{\rm p.c.}}=-i\sigma^\dagger\xi^{(n)}_{\lambda\mu}\sigma \nonumber \\*
  & P^{B1}_{\mu,2}  =   \sigma\xi^{(n)}_\mu\sigma\;,&\quad {\tilde P^{B1}_{\mu,2}} \equiv 
    (P^{B1}_{\mu,2})_{{\rm p.c.}}   =   \sigma^\dagger\xi^{(n)}_\mu\sigma^\dagger\nonumber \\*
  & P^{B1}_{\mu,3}  =   i\sigma\xi^{(n)}_{\mu5}\sigma\;,&\quad {\tilde P^{B1}_{\mu,3}} \equiv 
    (P^{B1}_{\mu,3})_{{\rm p.c.}}=-i\sigma^\dagger\xi^{(n)}_{5\mu}\sigma^\dagger\nonumber \\*
  & P^{B1}_{\mu,4}  =   i\sigma\xi^{(n)}_{\mu\lambda}\sigma^\dagger\;,&\quad {\tilde P^{B1}_{\mu,4}} \equiv 
    P^{B1}_{\mu,4} \;.
\end{eqnarray}
A comparison of \eq{L3B1O} with Eq.~(59) in \rcite{Aubin:StagHL2006}
shows that we have dropped the terms with coefficients $c^B_{16,k}$, $c^B_{17,k}$, 
and $c^B_{18,k}$ because one can write them as 
linear combinations of other terms in the Lagrangian using
\eqs{vslash-property1}{vslash-property5} below and the cyclic property of the trace.
For example, the term with coefficient $c^B_{16,k}$ is linearly dependent on the
terms with coefficients $c^B_{9,k}$ and $c^B_{13,k}$. Terms with coefficients
$c^B_{27,k}$ and $c^B_{28,k}$ in  Eq.~(60) of \rcite{Aubin:StagHL2006} have been dropped
in \eq{L3B1P} for the same reason.

For the type-B contributions to the current, we have:
\begin{eqnarray}\label{eq:j2B1}
j^{\mu,i \Xi}_{2,a^2,B1} & = & a^2\sum_{k=1}^3 
	     \sum_\nu \Biggl\{ r^{B1}_{5,k}\,
	      \trDt\left(\half T_\Xi\gamma^\mu (1-\gamma_5) H  
              v_\nu v^\nu \cO^{B1,+}_{\nu,k} \sigma^\dagger \lambda^{(i)}\right)\nonumber \\
        &&\hspace{2.2truecm} + r^{B1}_{6,k}\,
              \trDt\left(\half T_\Xi\gamma^\mu (1-\gamma_5) H \sigma^\dagger
	      \lambda^{(i)} \right)
	      v_\nu v^\nu \Tr(\cO^{B1,+}_{\nu,k}) \nonumber \\
	&&\hspace{2.2truecm} +
	      r^{B1}_{7,k}\,
	      \trDt\left(\half T_\Xi\gamma^\mu (1-\gamma_5) H  
	      v_\nu v^\nu \cO^{B1,-}_{\nu,k} \sigma^\dagger \lambda^{(i)}\right)\nonumber \\
	&&\hspace{2.2truecm} + r^{B1}_{8,k}\,
	      \trDt\left(\half T_\Xi\gamma^\mu (1-\gamma_5) H \sigma^\dagger
	      \lambda^{(i)}\right)
	      v_\nu v^\nu \Tr(\cO^{B1,-}_{\nu,k}) \nonumber \\
	&&\hspace{2.2truecm}  +r^{B1}_{9,k}\,
	      \trDt\left(\half T_\Xi\gamma^\mu (1-\gamma_5) H \gamma^\nu  
	      v_\nu  \cO^{B1,+}_{\nu,k} \sigma^\dagger \lambda^{(i)}\right)\nonumber \\
	&&\hspace{2.2truecm} + r^{B1}_{10,k}\,
	      \trDt\left(\half T_\Xi\gamma^\mu (1-\gamma_5) H \gamma^\nu \sigma^\dagger
	      \lambda^{(i)} \right)
	      v_\nu  \Tr(\cO^{B1,+}_{\nu,k}) \nonumber \\
	&&\hspace{2.2truecm} +
	      r^{B1}_{11,k}\,
	      \trDt\left(\half T_\Xi\gamma^\mu (1-\gamma_5) H \gamma^\nu  
	      v_\nu  \cO^{B1,-}_{\nu,k} \sigma^\dagger \lambda^{(i)}\right)\nonumber \\
	&&\hspace{2.2truecm} + r^{B1}_{12,k}\,
	      \trDt\left(\half T_\Xi\gamma^\mu (1-\gamma_5) H \gamma^\nu \sigma^\dagger
	      \lambda^{(i)}\right)
	      v_\nu  \Tr(\cO^{B1,-}_{\nu,k}) \Biggr\}\ .
\end{eqnarray}
Here we have omitted terms in \rcite{Aubin:StagHL2006} with coefficients
$r_{1,k}$ through $r_{4,k}$.  These terms have the (Lorentz and taste) index $\nu$ set to
$\mu$ and not summed over.  We believe such terms are inconsistent with 
heavy-quark spin symmetry, which is not broken by light-light four-quark operators in the SET.
In the next subsection, we give a more detailed 
discussion about type-B contributions to the current, which will further elucidate the reason for dropping
these terms.

\subsection{Discretization errors at NLO: Heavy-taste breaking terms}
\label{sec:NLOa2type2}

We now proceed to determine the chiral representatives of the light-heavy terms in
the SET, \eq{SET_lh}. The spin and taste matrices between
$\overline{Q}$ and $Q$ in this case mean that heavy-quark spin and taste symmetries
are broken. The corresponding chiral operators  are completely new, unrelated to those in
\rcite{Aubin:StagHL2006}, and we must determine them from scratch.  
That requires defining spurions to
make the operators ``invariant,'' and then constructing the possible chiral operators
in terms of those spurions. Initially, we do not allow additional derivatives
(\ie either the covariant derivative, $\rightvec D_\mu$ or the axial current 
$\mathbb{A}$), and find the
chiral operators summarized by the terms
$\cL_{2,a^2}^{A2}$, $\cL_{2,a^2}^{B2}$, 
$j^{\mu,i \Xi}_{2,a^2,A2}$ and $j^{\mu,i \Xi}_{2,a^2,B2}$ in \eqs{L2}{current2}. 
We then consider terms with a single additional derivative, which are summarized in
$\cL_{3,a^2}^{A2}$ and $\cL_{3,a^2}^{B2}$, \eq{L3}.

We take the type-A operator $\big[V\times P\big]^{lh}$ as an example:
\begin{eqnarray}
a^2\cO_{[V\times P]^{lh}} & \equiv & a^2 \ \overline{q}(\gamma^\mu\otimes\xi_5)q \;
                                \overline{Q}(\gamma_\mu\otimes\xi_5)Q ,\nonumber \\ 
                  & = & a^2[\overline{q}^L(\gamma^\mu\otimes\xi_5)q^L+\overline{q}^R(\gamma^\mu\otimes\xi_5)q^R] \;
                                [\overline{Q}(\gamma_\mu\otimes\xi_5)Q] ,\nonumber \\ 
                  & = & a^2[\overline{q}^L(\gamma^\mu\otimes A_1)q^L+\overline{q}^R(\gamma^{\mu}\otimes A_2)q^R] \;
                                [\overline{Q}\,\big(B(\mu)\otimes C\big)\, Q] \ , \label{eq:V*P-example}
\end{eqnarray}
with $q^L= [(1-\gamma_5)/2]\, q$ and $q^R= [(1+\gamma_5)/2]\, q$.
Note that \eq{V*P-example} is written in Minkowski space for consistency with 
the conventions of this paper. 
We have introduced four spurions, $A_1$, $A_2$, $B(\mu)$, and $C$, which 
transform as:  
\begin{eqnarray} 
A_1 & \rightarrow & L A_1 L^\dagger\;,\\
A_2 & \rightarrow & R A_2 R^\dagger\;,\\
B(\mu) & \rightarrow & S\,  B(\mu)\, S^\dagger\;, \\
C & \rightarrow & V  C V^\dagger \;.
\end{eqnarray}
Here $A_1$ and $A_2$ are light-quark spurions that transform according to the chiral
flavor-taste symmetry, while
$B(\mu)$ and $C$ transform to
maintain the spin and taste symmetry, respectively,  of the heavy quark. 
We will use them as building blocks for the chiral theory, and eventually let them
take the values
\begin{eqnarray}
A_1 & = & a \xi^{(n)}_5\equiv a\xi_5\otimes I_{\rm flavor}\;, \eqn{A1value}\\
A_2 & = & a \xi^{(n)}_5\equiv a\xi_5\otimes I_{\rm flavor}\;, \eqn{A2value}\\
B(\mu) & = & \gamma_\mu\;, \eqn{Bvalue}\\
C & = &  a\xi_5\;, \eqn{Cvalue}
\end{eqnarray}
where $I_{\rm flavor}$ is the identity in flavor space. We employ
two separate heavy quark spurions so that we can let $B(\mu)$ take its final value
before $A_1$, $A_2$, and $C$ do.
This two-stage procedure is useful in elucidating the implications of
Lorentz (or, equivalently, Euclidean rotation) invariance.  Since Lorentz transformations include
heavy-quark spin transformations, once $B(\mu)$ is introduced in the last
line of  \eq{V*P-example}, the 4-quark operator no longer transforms as a Lorentz scalar field.
The chiral operators we construct from $A_1$, $A_2$, and $B(\mu)\otimes C$ will thus be invariants
under heavy quark spin, heavy quark taste, and light quark chiral transformations, but not under
Lorentz transformations.  However, once we replace $B(\mu)$ by $\gamma_\mu$ 
(and sum over $\mu$),
the 4-quark operator is once again a Lorentz scalar, and so must be the resulting chiral
operators.

In constructing chiral operators from these spurions, we first note that $A_1$ and $A_2$
may be combined with $\sigma$ and $\sigma^\dagger$ in order to form objects
that transform with  $\mathbb{U}$ under the light-quark symmetries. This is convenient 
because $H$ and $\overline{H}$ transform
in that way, \eq{H-transform}.  We note
\begin{eqnarray} 
\sigma^\dagger A_1 \sigma&  \rightarrow &  \mathbb{U} \;(\sigma^\dagger A_1 \sigma)\; \mathbb{U}^\dagger  \;,\\
\sigma A_2 \sigma^\dagger&  \rightarrow  & \mathbb{U} \;(\sigma A_2 \sigma^\dagger)\;\mathbb{U}^\dagger  \;.
\end{eqnarray}
We can now easily make chiral operators that are invariant under heavy and light taste
symmetry and spin symmetry, and are
bilinear in $B(\mu)\otimes C$ and $A_1$ or $A_2$. (Terms with more spurions are higher order.)
We find the following operators:
\begin{eqnarray}
&\Tr\left[\overline{H} \,(B(\mu)\otimes C\big)\, H\,\Gamma_1 \sigma^\dagger A_1 \sigma \right], 
&\hspace{5mm} \Tr\left[\overline{H} \,(B(\mu)\otimes C\big)\, H\,\Gamma_2 \sigma A_2 \sigma^\dagger \right], \nonumber \\
&\Tr\left[\overline{H} \,(B(\mu)\otimes C\big)\, H\,\Gamma_3 \right] \Tr\left[ \sigma^\dagger A_1 \sigma \right], 
&\hspace{5mm}\Tr\left[\overline{H} \,(B(\mu)\otimes C\big)\, H\,\Gamma_4 \right] \Tr\left[\sigma A_2 \sigma^\dagger \right], 
\nonumber
\end{eqnarray}
where $\Gamma_1,\cdots,\Gamma_4$ are (for the moment, arbitrary) combinations of $\gamma$ matrices and components
of the heavy quark velocity $v$, which are the only additional factors allowed at this order.
Replacing $B(\mu)$ by $\gamma_\mu$, we may then demand Lorentz (and parity)
invariance.  The resulting operators are
\begin{eqnarray}
&&v^\mu \Tr\left(\overline{H} \gamma_\mu C H \sigma^\dagger A_1 \sigma \right) + 
v^\mu \Tr\left(\overline{H} \gamma_\mu C H \sigma A_2 \sigma^\dagger \right)\;, \label{eq:Hbar-B-H1a} \\
&&\Tr\left(\overline{H} \gamma_\mu C H \gamma^\mu \sigma^\dagger A_1 \sigma \right  ) + 
\Tr\left(\overline{H} \gamma_\mu C H \gamma^\mu \sigma A_2 \sigma^\dagger \right)\;, \label{eq:Hbar-B-H1b} \\
&&v^\mu \Tr\left(\overline{H} \gamma_\mu C H \right) \Tr\left( \sigma^\dagger A_1 \sigma \right) + 
v^\mu Tr\left(\overline{H} \gamma_\mu C H \right) \Tr\left(\sigma A_2 \sigma^\dagger \right)\;, \label{eq:Hbar-B-H1c} \\
&&\Tr\left(\overline{H}  \gamma_\mu C H \gamma^\mu \right) \Tr\left( \sigma^\dagger A_1 \sigma \right) + 
\Tr\left(\overline{H}  \gamma_\mu C H \gamma^\mu \right) \Tr\left(\sigma A_2 \sigma^\dagger \right)\ , \label{eq:Hbar-B-H1d}
\end{eqnarray}
Here, parity invariance requires that $A_1$ and $A_2$ enter symmetrically; there are
no parity-odd bilinears in $H$ and $\overline{H}$ that could be multiplied by an
antisymmetric combination of $A_1$ and $A_2$. We have also omitted the direct product symbol $\otimes$ where the meaning
is clear from context.
Since
$ \Tr\left( \sigma^\dagger A_1 \sigma \right) =  0 = \Tr\left( \sigma^\dagger A_2 \sigma \right) $
once $A_1$ and $A_2$ take their final values, \eq{Hbar-B-H1c} 
and \eq{Hbar-B-H1d} may be dropped. 
On the other hand, various simplifications of terms involving  $H$ and $\overline{H}$ 
are possible here and below, due to the overall factors of $(1+\vslash)$ in their 
definitions [\eqs{H_definition}{Hbar_definition}], the fact that $v^2=1$, and the relation
$v^\mu B^*_{\mu\alpha a}=0$ for the vector meson field $B^*$.
We list some relations that 
are useful for simplifying terms:
\begin{eqnarray}
&& \vslash \Bslash^* = - \Bslash^* \vslash\quad\Rightarrow\quad \vslash H = - H \vslash, \label{eq:vslash-property1} \\ 
&&(1+\vslash) \vslash = (1+\vslash), \label{eq:vslash-property2} \\
&&(1+\vslash)\gamma_5 (1+\vslash) =  0,\label{eq:vslash-property3} \\
&&(1+\vslash)\gamma_\mu (1+\vslash) =  (1+\vslash) v_\mu (1+\vslash),\label{eq:vslash-property4}\\
&&(1-\vslash)\gamma_{\mu\nu} (1+\vslash) =  (1-\vslash) \left(\gamma_\mu v_\nu -\gamma_\nu v_\mu\right)  (1+\vslash),\label{eq:vslash-property5}\\
&&\trD(\overline{H}H\gamma_\mu) =  -v_\mu \trD(\overline{H}H),\label{eq:vslash-property6}
\end{eqnarray}
where $\trD$ is a trace over Dirac indices only, and \eq{vslash-property6} is actually a simple consequence of
\eqs{vslash-property1}{vslash-property4} and the cyclic property of the trace.
With these relations, it is straightforward
show that \eq{Hbar-B-H1a} and \eq{Hbar-B-H1b} are both proportional
to  
\begin{eqnarray}
a^2 \Tr\left(\overline{H} \xi_5 H \sigma^\dagger \xi^{(n)}_5 \sigma \right) + 
a^2 \Tr\left(\overline{H} \xi_5 H \sigma \xi^{(n)}_5 \sigma^\dagger \right)\ ,
\end{eqnarray}
where we wave inserted final values of the spurions from \eqsthree{A1value}{A2value}{Cvalue}.
We then follow the same procedure for other type-A operators. For clarity, we write the terms 
with a single trace and terms with two traces separately. First, we list the single-trace terms:
\begin{eqnarray}\eqn{SAsingle}
\big[S\times A\big] & \rightarrow & a^2 \Tr\left(\overline{H} \xi_{5\mu} H \sigma^\dagger \xi^{(n)}_{\mu5} \sigma^\dagger \right) + 
                                    a^2 \Tr\left(\overline{H} \xi_{5\mu} H \sigma \xi^{(n)}_{\mu5} \sigma \right)\;, \eqn{SxA}\\
\big[S\times V\big] & \rightarrow & a^2 \Tr\left(\overline{H} \xi_{\mu} H \sigma^\dagger \xi^{(n)}_{\mu} \sigma^\dagger \right) + 
                                    a^2 \Tr\left(\overline{H} \xi_{\mu} H \sigma \xi^{(n)}_{\mu} \sigma \right)\;,\\
\big[P\times A\big] & \rightarrow & 0\;, \eqn{PxA} \\
\big[P\times V\big] & \rightarrow & 0\;,  \\
\big[T\times A\big] & \rightarrow & a^2 \Tr\left(\overline{H} \gamma_{\lambda\nu} \xi_{5\mu} H \gamma^{\nu\lambda} \sigma^\dagger \xi^{(n)}_{\mu5} \sigma^\dagger \right) + 
				     a^2 \Tr\left(\overline{H} \gamma_{\lambda\nu} \xi_{5\mu} H \gamma^{\nu\lambda} \sigma \xi^{(n)}_{\mu5} \sigma \right)\;, \eqn{TxA}\\
\big[T\times V\big] & \rightarrow & a^2 \Tr\left(\overline{H} \gamma_{\lambda\nu} \xi_{\mu}  H \gamma^{\nu\lambda} \sigma^\dagger \xi^{(n)}_{\mu} \sigma^\dagger \right) + 
				     a^2 \Tr\left(\overline{H} \gamma_{\lambda\nu} \xi_{\mu}  H \gamma^{\nu\lambda} \sigma \xi^{(n)}_{\mu} \sigma \right)\;,\\
\big[V\times S\big] & \rightarrow & a^2 \Tr\left(\overline{H} H \right)\;,\\
\big[V\times P\big] & \rightarrow & a^2 \Tr\left(\overline{H} \xi_5 H \sigma^\dagger \xi^{(n)}_5 \sigma \right) + 
				     a^2 \Tr\left(\overline{H} \xi_5 H \sigma \xi^{(n)}_5 \sigma^\dagger \right)\;,\\
\big[V\times T\big] & \rightarrow & a^2 \Tr\left(\overline{H} \xi_{\nu\lambda} H \sigma^\dagger \xi^{(n)}_{\lambda\nu} \sigma \right) + 
				     a^2 \Tr\left(\overline{H} \xi_{\nu\lambda} H \sigma \xi^{(n)}_{\lambda\nu} \sigma^\dagger \right)\;,\\
\big[A\times S\big] & \rightarrow & a^2 \Tr\left(\overline{H} \gamma_{5\mu} H \gamma^{\mu 5} \right)\;,\\
\big[A\times P\big] & \rightarrow & a^2 \Tr\left(\overline{H} \gamma_{5\mu} \xi_5 H \gamma^{\mu 5} \sigma^\dagger \xi^{(n)}_5 \sigma \right) + 
				     a^2 \Tr\left(\overline{H} \gamma_{5\mu} \xi_5 H \gamma^{\mu 5} \sigma \xi^{(n)}_5 \sigma^\dagger \right)\;,\\
\big[A\times T\big] & \rightarrow & a^2 \Tr\left(\overline{H} \gamma_{5\mu} \xi_{\nu\lambda} H \gamma^{\mu 5} \sigma^\dagger \xi^{(n)}_{\lambda\nu} \sigma \right) + 
				     a^2 \Tr\left(\overline{H} \gamma_{5\mu} \xi_{\nu\lambda} H \gamma^{\mu 5} \sigma \xi^{(n)}_{\lambda\nu} \sigma^\dagger \right) \ .
\end{eqnarray}
As before, all twice-repeated indices are summed.
The double-trace terms are:
\begin{eqnarray}
\big[S\times A\big] & \rightarrow & a^2 \Tr\left(\overline{H} \xi_{5\mu} H \right)
					\Tr\left(\sigma^\dagger \xi^{(n)}_{\mu5} \sigma^\dagger \right) + 
                                    a^2 \Tr\left(\overline{H} \xi_{5\mu} H \right)
					\Tr\left(\sigma \xi^{(n)}_{\mu5} \sigma \right)\;,\\
\big[S\times V\big] & \rightarrow & a^2 \Tr\left(\overline{H} \xi_{\mu} H \right)
					\Tr\left(\sigma^\dagger \xi^{(n)}_{\mu} \sigma^\dagger \right) + 
                                    a^2 \Tr\left(\overline{H} \xi_{\mu} H \right)
					\Tr\left(\sigma \xi^{(n)}_{\mu} \sigma \right)\;,\\
\big[P\times A\big] & \rightarrow & 0\;,\\
\big[P\times V\big] & \rightarrow & 0\;,\\
\big[T\times A\big] & \rightarrow & a^2 \Tr\left(\overline{H} \gamma_{\lambda\nu} \xi_{5\mu} H \gamma^{\lambda\nu} \right)
					\Tr\left(\sigma^\dagger \xi^{(n)}_{\mu5} \sigma^\dagger \right) + 
                                    a^2 \Tr\left(\overline{H} \gamma_{\lambda\nu} \xi_{5\mu} H \gamma^{\lambda\nu} \right)
					\Tr\left(\sigma \xi^{(n)}_{\mu5} \sigma \right)\;,\\
\big[T\times V\big] & \rightarrow & a^2 \Tr\left(\overline{H} \gamma_{\lambda\nu} \xi_{\mu}  H \gamma^{\lambda\nu} \right)
					\Tr\left(\sigma^\dagger \xi^{(n)}_{\mu} \sigma^\dagger \right) + 
                                    a^2 \Tr\left(\overline{H} \gamma_{\lambda\nu} \xi_{\mu}  H \gamma^{\lambda\nu} \right)
					\Tr\left(\sigma \xi^{(n)}_{\mu} \sigma \right)\;,\\
\big[V\times S\big] & \rightarrow & 0\;,\\
\big[V\times P\big] & \rightarrow & 0\;,\\
\big[V\times T\big] & \rightarrow & 0\;\\
\big[A\times S\big] & \rightarrow & 0\;,\\
\big[A\times P\big] & \rightarrow & 0\;,\\
\big[A\times T\big] & \rightarrow & 0\;.\eqn{ATdouble}
\end{eqnarray}
In \eqsthru{SAsingle}{ATdouble}, we have again used the fact that Lorentz-invariant,
parity-odd bilinears in
$H$ and $\overline{H}$ [such as $\trD(\overline{H}H\gamma_5)$, 
$\trD(\overline{H}\gamma^\mu H\gamma_{\mu5})$, or 
$\trD(\overline{H}\gamma^{\mu\nu} H\gamma_{\nu\mu}\gamma_5)$]
vanish.  The reason for this is 
that, once the Dirac traces are performed, the only objects from which to form
invariants in the heavy-meson sector are $B$, $B^\dagger$, $B^*_\mu$, 
$B^{\dagger*}_\nu$, and $v_\lambda$, and it is not possible to make a Lorentz-invariant
bilinear in the meson fields that is parity odd out of these ingredients. This eliminates the
possibility of antisymmetric combinations of the light quark spurions, multiplied by
parity-odd combinations of the heavy-meson fields.

We now consider the type-B operators.  The procedure here is a bit more complicated because
these operators violate Lorentz invariance in a particular way,
and we must ensure that the chiral operators do the same.  Our approach is based on 
that introduced by Sharpe and
Van de Water \cite{Sharpe:SChiPT2005} to find light-meson chiral representatives
of type-B operators.
We take  $\big[T_\mu\times A_\mu \big]$ as an example:
\begin{equation}
 a^2\cO_{[T_\mu\times A_\mu]} \equiv  a^2\sum_\mu \Biggl\{
      \overline{q}_l(\gamma^{\mu\nu}\otimes\xi_{\mu5})q_l \overline{q}_h(\gamma_{\nu\mu}\otimes\xi_{5\mu})q_h
     -\overline{q}_l(\gamma^{\mu\nu5}\otimes\xi_{\mu5})q_l \overline{q}_h(\gamma_{5\nu\mu}\otimes\xi_{5\mu})q_h
                \Biggr\}\;.
\end{equation}
The second term in this expression removes the Lorentz-singlet component. However, it is
unnecessary to keep both terms here because 
the second term can be written as a linear combination 
of the first term and $\big[T\times A \big]$, which has already have been taken
into account. Further, it is useful for the moment to remove the sums (explicit or implicit)
over the indices $\mu,\nu$.  Thus we are led to consider the operator
\begin{eqnarray}\label{eq:T-mu*A-mu}
a^2\cO(\mu,\nu)& \equiv & a^2\overline{q}(\gamma^{\mu\nu}\otimes\xi_{\mu5})q \ 
				          \overline{Q}(\gamma_{\nu\mu}\otimes\xi_{5\mu})Q  ,\nonumber \\ 
			     &  =   & a^2 \Bigl[\overline{q}^L(\gamma^{\mu\nu}\otimes\xi_{\mu5})q^R
                                         +\overline{q}^R(\gamma^{\mu\nu}\otimes\xi_{\mu5})q^L\Bigr] 
                                         \Bigl[\overline{Q}(\gamma_{\nu\mu}\otimes\xi_{5\mu})Q\Bigr]  ,\nonumber \\ 
			     &  =   & \Bigl[\overline{q}^L\big(\gamma^{\mu\nu}\otimes A_1(\mu)\big)q^R
					  +\overline{q}^R\big(\gamma^{\mu\nu}\otimes A_2(\mu)\big) q^L\Bigr] 
                                         \Bigl[\overline{Q} \big(B(\nu,\mu)\otimes C(\mu)\big) Q\Bigr] ,
\end{eqnarray}
where $\mu$ and $\nu$ are fixed.
With the spurions $A_1(\mu)$, $A_2(\mu)$, and $B(\nu,\mu)\otimes C(\mu)$, 
we can construct two single-trace $\cO(a^2)$ terms that are invariant
under heavy and light taste symmetry and heavy-quark spin symmetry:
\begin{eqnarray}
&&\Tr\left[\overline{H} \big(B(\nu,\mu)\otimes C(\mu)\big) H \Gamma_1\,  \sigma^\dagger A_1(\mu) \sigma^\dagger \right],\nonumber \\
&&\Tr\left[\overline{H} \big(B(\nu,\mu)\otimes C(\mu)\big) H \Gamma_2\,  \sigma A_2(\mu)  \sigma \right],
\eqn{Bpossible1}
\end{eqnarray}
where $\Gamma_1$ and $\Gamma_2$ are as-yet undetermined 
combinations of $\gamma$ matrices and components of $v$.  There are also two-trace
versions of these operators, in which the heavy- and light-quark factors are separately traced, but 
for simplicity we focus on the single-trace case here.

We now replace the spurion $B(\nu,\mu)$ with its value $\gamma_{\nu\mu}$. 
We also restore the sum over $\nu$ (but not $\mu$), 
considering chiral representatives of the 
operator $a^2\cO(\mu) = a^2 \sum_\nu\cO(\mu,\nu)$:
\begin{eqnarray}\label{eq:T-mu*A-mu}
a^2\cO(\mu)& = & \Bigl[\overline{q}^L\gamma^{\mu\nu} A_1(\mu)q^R
+\overline{q}^R\gamma^{\mu\nu} A_2(\mu) q^L\Bigr] 
\Bigl[\overline{Q} \gamma_{\nu\mu} C(\mu) Q\Bigr]  \hspace{5mm} (\mu\ {\rm fixed})\ ,
\end{eqnarray}
with the $\otimes$ symbols and the sum on $\nu$ implicit. 
The operator $\cO(\mu)$ is the $\;{}^\mu{}_\mu$ component
of a two-index Lorentz tensor, and is therefore a linear combination of an element
of a symmetric traceless tensor and 
a Lorentz singlet (the trace). The singlet piece, in which the sum over the Lorentz index
$\mu$ is decoupled from the taste label $\mu$ of the spurions, is simply a repeat of the
corresponding type-A operator; only the symmetric tensor is new. 
Thus the desired chiral operators
are $\;{}^\mu{}_\mu$ components of two-index Lorentz tensors, where it is not necessary
to insist on tracelessness because the trace term again will repeat one of
the type-A chiral operators. From the possibilities in \eq{Bpossible1}, two independent 
operators may now be constructed:
\begin{eqnarray}
&&\Tr\left[\overline{H} \gamma_{\nu\mu} C(\mu) H \gamma^{\mu\nu} \left(
\sigma^\dagger A_1(\mu) \sigma^\dagger   + 
\sigma A_2(\mu)  \sigma \right)\right]\;,
\nonumber \\
&& \Tr\left[\overline{H} \gamma_{\nu\mu} C(\mu)
H \gamma^{\mu\nu}\gamma_5 \left( \sigma^\dagger A_1(\mu) \sigma^\dagger  - 
\sigma A_2(\mu)\sigma \right)\right] \;, \eqn{Bpossible2}
\end{eqnarray}
with $\mu$  still fixed.  Using \eqsthru{vslash-property1}{vslash-property5}, it is not
hard to show that choices other than $\gamma^{\mu\nu}$ for the $\Gamma_i$ factors
following the $H$ field either vanish identically (\eg for the choice $\gamma^\mu v^{\nu}$) or
are proportional to one of the terms listed (\eg for the choice  $v^\mu \gamma^{\nu}$).
The symmetric  combination of $A_1$ and $A_2$ in the first term, as
well as the antisymmetric combination in the second, are required by parity.

Finally, we put in the fixed values of the spurions $A_1(\mu)$, $A_2(\mu)$, and $C(\mu)$,
and restore the sum on $\mu$, giving the two operators
\begin{eqnarray}
 && a^2 \sum_\mu \left\{ \Tr\left[\overline{H} \gamma_{\nu\mu} \xi_{5\mu} H \gamma^{\mu\nu} 
\left(\sigma^\dagger \xi^{(n)}_{\mu5} \sigma^\dagger + 
\sigma \xi^{(n)}_{\mu5} \sigma \right)\right] \right\}\;,\\ \eqn{even-even_TmuxAmu}
 &&a^2 \sum_\mu \left\{ \Tr\left[\overline{H} \gamma_{\nu\mu} \xi_{5\mu} 
H \gamma^{\mu\nu}\gamma_5 \left( \sigma^\dagger \xi^{(n)}_{\mu5} \sigma^\dagger  - 
\sigma \xi^{(n)}_{\mu5} \sigma \right)\right] \right\} \;. \eqn{odd-odd_TmuxAmu}
\end{eqnarray}
As mentioned earlier, terms like \eq{odd-odd_TmuxAmu} (odd in the light spurions) are ruled
out in the type-A case 
by parity and Lorentz invariance. Here, however, Lorentz invariance is broken,
and $\trD\left(\overline{H} \gamma_{\nu\mu}H
\gamma^{\mu\nu}\gamma_5\right)$ does not vanish since the sum on $\mu$ is not free, but
coupled to the taste sum.  Further, one can check that the term in \eq{odd-odd_TmuxAmu}
is Hermitian and time-reversal invariant; 
for details of how time-reversal
symmetry acts on relevant quantities, see Ref.~\cite{Aubin:StagHL2006}, Sec.~III D.

We derive the other type-B terms similarly. For clarity, we 
write the single-trace terms and 
the double-trace terms separately. The single-trace terms are:
\begin{eqnarray}
\big[T_\mu\times A_\mu \big] & \rightarrow & 
         a^2 \sum_\mu \left\{ \Tr\left[\overline{H} \gamma_{\nu\mu} \xi_{5\mu} H \gamma^{\mu\nu} 
\left(\sigma^\dagger \xi^{(n)}_{\mu5} \sigma^\dagger  + 
	  \sigma \xi^{(n)}_{\mu5} \sigma \right)\right] \right\}\;,\nonumber\\
         &&a^2 \sum_\mu \left\{ \Tr\left[\overline{H} \gamma_{\nu\mu} \xi_{5\mu} H \gamma^{\mu\nu} \gamma_5
\left(\sigma^\dagger \xi^{(n)}_{\mu5} \sigma^\dagger  - 
	  \sigma \xi^{(n)}_{\mu5} \sigma \right)\right] \right\}\;;\eqn{TmuxAmu}\\
	  \big[T_\mu\times V_\mu \big] & \rightarrow & 
         a^2 \sum_\mu \left\{ \Tr\left[\overline{H} \gamma_{\nu\mu} \xi_{\mu}  H \gamma^{\mu\nu} 
\left(\sigma^\dagger \xi^{(n)}_{\mu } \sigma^\dagger + 
\sigma \xi^{(n)}_{\mu } \sigma \right) \right]\right\}\;,\nonumber\\
        && a^2 \sum_\mu \left\{ \Tr\left[\overline{H} \gamma_{\nu\mu} \xi_{\mu}  H \gamma^{\mu\nu} \gamma_5
\left(\sigma^\dagger \xi^{(n)}_{\mu } \sigma^\dagger - 
\sigma \xi^{(n)}_{\mu } \sigma \right) \right]\right\}\;;\\
\big[A_\mu\times T_\mu \big] & \rightarrow & 
         a^2 \sum_\mu \left\{ \Tr\left[\overline{H} \gamma_{5\mu} \xi_{\nu\mu} H \gamma^{\mu 5} 
\left(\sigma^\dagger \xi^{(n)}_{\mu\nu} \sigma  + 
	  \sigma \xi^{(n)}_{\mu\nu} \sigma^\dagger \right)\right] \right\}\;,\nonumber\\
         &&a^2 \sum_\mu \left\{ \Tr\left[\overline{H} \gamma_{5\mu} \xi_{\nu\mu} H \gamma^{\mu} 
\left(\sigma^\dagger \xi^{(n)}_{\mu\nu} \sigma  - 
	  \sigma \xi^{(n)}_{\mu\nu} \sigma^\dagger \right)\right] \right\}\;;\\
\big[V_\mu\times T_\mu \big] & \rightarrow & 
         a^2 \sum_\mu \left\{ v^\mu v_{\mu} \Tr\left[\overline{H} \xi_{\nu\mu} H 
\left(\sigma^\dagger \xi^{(n)}_{\mu\nu} \sigma      + 
	  \sigma         \xi^{(n)}_{\mu\nu} \sigma^\dagger \right)\right] \right\}\;, \nonumber \\
         &&a^2 \sum_\mu \left\{ v^\mu \Tr\left[\overline{H} \xi_{\nu\mu} H \gamma_{\mu5}
\left(\sigma^\dagger \xi^{(n)}_{\mu\nu} \sigma      - 
	  \sigma         \xi^{(n)}_{\mu\nu} \sigma^\dagger \right)\right] \right\} \ . 
\end{eqnarray}
The double-trace terms are:
\begin{eqnarray}
\big[T_\mu\times A_\mu \big] & \rightarrow & 
         a^2 \sum_\mu \Tr\left(\overline{H} \gamma_{\nu\mu} \xi_{5\mu} H \gamma^{\mu\nu} \right) 
                 \left\{ \Tr\left(\sigma^\dagger \xi^{(n)}_{\mu5} \sigma^\dagger \right) + 
                         \Tr\left(\sigma         \xi^{(n)}_{\mu5} \sigma         \right) \right\}\;,\nonumber\\
         &&a^2 \sum_\mu \Tr\left(\overline{H} \gamma_{\nu\mu} \xi_{5\mu} H \gamma^{\mu\nu}\gamma_5 \right) 
                 \left\{ \Tr\left(\sigma^\dagger \xi^{(n)}_{\mu5} \sigma^\dagger \right) - 
                         \Tr\left(\sigma         \xi^{(n)}_{\mu5} \sigma         \right) \right\}\;;\\
\big[T_\mu\times V_\mu \big] & \rightarrow & 
         a^2 \sum_\mu \Tr\left(\overline{H} \gamma_{\nu\mu} \xi_{\mu} H \gamma^{\mu\nu} \right) 
                 \left\{ \Tr\left(\sigma^\dagger \xi^{(n)}_{\mu} \sigma^\dagger \right) + 
                         \Tr\left(\sigma         \xi^{(n)}_{\mu} \sigma         \right) \right\}\;,\nonumber\\
         &&a^2 \sum_\mu \Tr\left(\overline{H} \gamma_{\nu\mu} \xi_{\mu} H \gamma^{\mu\nu}
\gamma_5 \right) 
                 \left\{ \Tr\left(\sigma^\dagger \xi^{(n)}_{\mu} \sigma^\dagger \right) - 
                         \Tr\left(\sigma         \xi^{(n)}_{\mu} \sigma         \right) \right\}\;;\\
\big[A_\mu\times T_\mu \big] & \rightarrow & 0\;,\\
\big[V_\mu\times T_\mu \big] & \rightarrow & 0 \ .
\end{eqnarray}

The Lagrangian terms $\cL^{A2}_{2,a^2}$ and $\cL^{B2}_{2,a^2}$, 
\eq{L2}, collect the (heavy-quark taste violating) chiral operators that we have derived so far.
To make the notation a bit more compact, we first define the operators:
\begin{eqnarray}\label{eq:A2}
  P^{\pm}_{5} & = & \frac{1}{2} (\sigma\xi^{(n)}_5\sigma^{\dagger} \pm {\rm p.c.}) ,\nonumber \\
  P^{\pm}_{\mu \nu} & = & \frac{1}{2}  (\sigma\xi^{(n)}_{\mu \nu}\sigma^{\dagger} \pm {\rm p.c.}) ,\nonumber \\
  P^{\pm}_{\mu} & = &  \frac{1}{2} (\sigma\xi^{(n)}_\mu\sigma \pm {\rm p.c.}) ,\nonumber \\ 
  P^{\pm}_{\mu 5} & = & \frac{1}{2}  (\sigma\xi^{(n)}_{\mu 5}\sigma \pm {\rm p.c.})\ .
\end{eqnarray}
We then have 
\begin{eqnarray}
\cL^{A2}_{2,a^2} & = &   a^2\Biggl\{ 
               K^{A2}_{1,0} \Tr\left(\overline{H} H\right)
	      +K^{A2}_{1,1} \Tr\left(\overline{H} \xi_{5}    H P^{+}_{5} \right) +
	      K^{A2}_{1,2} \Tr\left(\overline{H} \xi_{\mu}  H P^{+}_{\mu} \right) \nonumber \\*&&{}+
	      K^{A2}_{1,3} \Tr\left(\overline{H} \xi_{5\mu} H P^{+}_{\mu 5} \right) +
	      K^{A2}_{1,4} \Tr\left(\overline{H} \xi_{\mu\nu} H P^{+}_{\nu\mu} \right) +
	      K^{A2}_{1,5} \Tr\left(\overline{H} \gamma_{5\mu} H  \gamma^{\mu 5} \right) \nonumber \\*&&{}+
	      K^{A2}_{1,6} \Tr\left(\overline{H} \gamma_{5\mu} \xi_{5 } H  \gamma^{\mu 5} P^{+}_{5} \right)  +
	      K^{A2}_{1,7} \Tr\left(\overline{H} \gamma_{\mu\nu} \xi_{\lambda } H  \gamma^{\nu\mu} P^{+}_{\lambda} \right) \nonumber \\*&&{}+
	      K^{A2}_{1,8} \Tr\left(\overline{H} \gamma_{\mu\nu} \xi_{5\lambda} H  \gamma^{\nu\mu} P^{+}_{\lambda 5} \right) +
	      K^{A2}_{1,9} \Tr\left(\overline{H} \gamma_{5\mu} \xi_{\nu\lambda} H  \gamma^{\mu 5} P^{+}_{\lambda\nu} \right) \label{eq:L2A2} \\*&&{}+
	      K^{A2}_{2,1} \Tr\left(\overline{H} \xi_{ \mu} H \right) \Tr\left( P^{+}_{\mu  } \right) +
	      K^{A2}_{2,2} \Tr\left(\overline{H} \xi_{5\mu} H \right) \Tr\left( P^{+}_{\mu 5} \right) \nonumber \\*&&{}+
	      K^{A2}_{2,3} \Tr\left(\overline{H} \gamma_{\mu\nu} \xi_{\lambda } H  \gamma^{\nu\mu} \right) \Tr\left( P^{+}_{\lambda} \right) +
	      K^{A2}_{2,4} \Tr\left(\overline{H} \gamma_{\mu\nu} \xi_{5\lambda} H  \gamma^{\nu\mu} \right) \Tr\left( P^{+}_{\lambda 5} \right) 
			     \Biggr\} \ , \nonumber 
\end{eqnarray}
where ten terms are single-trace and four are double-trace.  
For completeness we have kept the trivial term $a^2K^{A2}_{1,0} \Tr\left(\overline{H} H\right)$ even though
it does not break any symmetries and just gives equal mass shifts to all tastes of pseudoscalar and vector heavy-light mesons. 
In fact, this term also would appear in  $\cL^{A1}_{2,a^2}$ but was dropped from 
Ref.~\cite{Aubin:StagHL2006} due to its triviality.
It is worth mentioning that the terms breaking the spin symmetry by $\gamma_{\mu\nu}$ in \eq{L2A2} 
can be replaced with simpler terms using the following identity:
\begin{equation}
 \overline{H} \gamma_{\mu\nu} T_\Xi H  \gamma^{\nu\mu}
 =\overline{H}\gamma_5 \gamma_{\mu\nu} T_\Xi H \gamma_5 \gamma^{\nu\mu}
 = - 2 \overline{H} \gamma_{5\rho} T_\Xi H  \gamma^{\rho5},
\end{equation}
where the first equality follows from the fact that the $\gamma_5$ factors just
interchange the components of
$\gamma_{\mu\nu}$, and the second 
can be proved using \eqsthree{vslash-property2}{vslash-property3}{vslash-property5}.

For $\cL^{B2}_{2,a^2}$ we have
\begin{eqnarray}
\cL^{B2}_{2,a^2} & = &   a^2 \sum_\mu \Biggl\{
	      K^{B2}_{1,1} \Tr\left(\overline{H} \gamma_{\nu\mu} \xi_{\mu } H  \gamma^{\mu\nu} P^{+}_{\mu  } \right) +
	      K^{B2}_{1,2} \Tr\left(\overline{H} \gamma_{\nu\mu} \xi_{5\mu} H  \gamma^{\mu\nu} P^{+}_{\mu 5} \right)  \nonumber \\*&&{}+
	      K^{B2}_{1,3} v^\mu v_\mu \Tr\left(\overline{H}   \xi_{\nu\mu} H  P^{+}_{\mu\nu} \right) +
	      K^{B2}_{1,4} \Tr\left(\overline{H} \gamma_{5\mu} \xi_{\nu\mu} H  \gamma^{\mu 5} P^{+}_{\mu\nu} \right)  \nonumber \\*&&{}+
	      K^{B2}_{1,5} \Tr\left(\overline{H} \gamma_{\nu\mu} \xi_{\mu } H  \gamma^{\mu\nu}\gamma_5 P^{-}_{\mu  } \right) + 
	      K^{B2}_{1,6} \Tr\left(\overline{H} \gamma_{\nu\mu} \xi_{5\mu} H  \gamma^{\mu\nu}\gamma_5 P^{-}_{\mu 5} \right)  \nonumber \\*&&{}+
	      K^{B2}_{1,7} v^\mu  \Tr\left(\overline{H}   \xi_{\nu\mu} H\gamma_{\mu5}  P^{-}_{\mu\nu} \right) +
	      K^{B2}_{1,8} v^\mu  \Tr\left(\overline{H} \gamma_{5\mu} \xi_{\nu\mu} H   P^{-}_{\mu\nu} \right) \eqn{L2B2}  \\*&&{}+
	      K^{B2}_{2,1} \Tr\left(\overline{H} \gamma_{\nu\mu} \xi_{\mu } H  \gamma^{\mu\nu} \right) \Tr\left( P^{+}_{\mu  } \right) +
	      K^{B2}_{2,2} \Tr\left(\overline{H} \gamma_{\nu\mu} \xi_{5\mu} H  \gamma^{\mu\nu} \right) \Tr\left( P^{+}_{\mu 5} \right) \nonumber \\*&&{}+
	      K^{B2}_{2,3} \Tr\left(\overline{H} \gamma_{\nu\mu} \xi_{\mu } H  \gamma^{\mu\nu}\gamma_5 \right) \Tr\left( P^{-}_{\mu  } \right) +
	      K^{B2}_{2,4} \Tr\left(\overline{H} \gamma_{\nu\mu} \xi_{5\mu} H  \gamma^{\mu\nu}\gamma_5 \right) \Tr\left( P^{-}_{\mu 5} \right) 
	      		       \Biggr\} \ . \nonumber 
\end{eqnarray}
where eight terms are single-trace and four are double-trace. 

There are a large number of terms contributing to the remaining NLO parts of the Lagrangian,  $\cL^{A2}_{3,a^2}$ and $\cL^{B2}_{3,a^2}$. An extra derivative, either in the form of the
covariant derivative $ D_\nu$ or the axial current $\mathbb{A}_\nu$, can be added to
the terms in $\cL^{A2}_{2,a^2}$ and $\cL^{B2}_{2,a^2}$ in many ways when one takes into 
account the ordering of terms and the various possibilities for contracting indices.  Faced with
this explosion of terms, we content ourselves with listing some representative contributions.
For all practical applications at NLO that we can envision, this will be sufficient, since in a lattice 
computation of some physical quantity one is only interested in knowing what analytic terms
are possible, and whether the coefficients of these terms are linearly dependent or independent,
 and not in knowing how to write those coefficients as combinations of the low energy constants in 
 the chiral Lagrangian.  This is the case for the heavy-light decay constant, discussed in \secref{fD}.  
For the NLO taste splittings of the masses of heavy-light mesons, treated \secref{Mass}, the
 quantities $\cL^{A2}_{3,a^2}$ and $\cL^{B2}_{3,a^2}$ are in fact irrelevant, because they either
have an extra factor of the residual momentum $k$, which vanishes on shell at this order, or 
because they have  an extra pion field at tree level.

Some representative contributions to $\cL^{A2}_{3,a^2}$ are: 
\begin{eqnarray}
\cL^{A2}_{3,a^2} & = &   a^2\Biggl\{ 
      \biggl[ i c^{A2}_{1,0} \Tr\left(\overline{H}  H v\negcdot \leftvec D
			      - v\negcdot \rightvec D \, \overline{H} H\right) +    \cdots  \nonumber \\*&&{}+
	      i c^{A2}_{1,9} \Tr\left(\overline{H} \gamma_{5\mu} \xi_{\nu\lambda} H \gamma^{\mu 5} v\negcdot \leftvec D\, P^{+}_{\lambda\nu}
			      - v\negcdot \rightvec D \, \overline{H} \gamma_{5\mu} \xi_{\nu\lambda} H \gamma^{\mu 5} P^{+}_{\lambda\nu}\right) 
      \biggr] \nonumber \\*&&{}+
      \biggl[ i c^{A2}_{2,1} \Tr\left(\overline{H} \xi_{\mu} H v\negcdot \leftvec D\, 
			      - v\negcdot \rightvec D \, \overline{H} \xi_{\mu} H \right) \Tr\left( P^{+}_{\mu}\right) + \cdots \nonumber \\*&&{}+
	      i c^{A2}_{2,4} \Tr\left(\overline{H} \gamma_{\mu\nu} \xi_{5\lambda} H \gamma^{\nu\mu} v\negcdot \leftvec D\, 
			      - v\negcdot \rightvec D \, \overline{H} \gamma_{\mu\nu} \xi_{5\lambda} H \gamma^{\nu\mu}\right) \Tr\left( P^{+}_{\lambda 5}\right) 
      \biggr] \nonumber \\*&&{}+
      \biggl[   c^{A2}_{3,0} \Tr\left( \overline{H} H \gamma_\sigma\gamma_5\mathbb{A}^\sigma\right) + \cdots +
	      c^{A2}_{3,9} \Tr\left( \overline{H} \gamma_{5\mu} \xi_{\nu\lambda} H \gamma^{\mu 5} \gamma_\sigma\gamma_5\{\mathbb{A}^\sigma, P^{+}_{\lambda\nu}\}\right)
      \biggr] \nonumber \\*&&{}+
      \biggl[ c^{A2}_{4,1} \Tr\left( \overline{H} \xi_{\mu} H \gamma_\sigma\gamma_5\mathbb{A}^\sigma \right) \Tr\left( P^{+}_{\mu}\right) + \cdots +
	      c^{A2}_{4,4} \Tr\left( \overline{H} \gamma_{\mu\nu} \xi_{5\lambda} H \gamma^{\nu\mu} \gamma_\sigma\gamma_5\mathbb{A}^\sigma \right) \Tr\left( P^{+}_{\lambda 5}\right) 
      \biggr]  \nonumber \\*&&{}+
      \biggl[ c^{A2}_{5,1} \Tr\left( \overline{H} \xi_{5} H \gamma_\sigma \gamma_5 \right) \Tr\left( \mathbb{A}^\sigma  P^{+}_{5}\right)  + \cdots +
	      c^{A2}_{5,9} \Tr\left( \overline{H} \gamma_{5\mu} \xi_{\nu\lambda} H \gamma^{\mu 5}  \gamma_\sigma\gamma_5 \right) \Tr\left( \mathbb{A}^\sigma P^{+}_{\lambda\nu}\right)
      \biggr]  \nonumber \\*&&{}+
      \biggl[ c^{A2}_{6,1} \Tr\left( \overline{H} \xi_{5} H \gamma_\sigma\gamma_5 [\mathbb{A}^\sigma, P^{-}_{5}] \right) + \cdots +
	      c^{A2}_{6,9} \Tr\left( \overline{H} \gamma_{5\mu} \xi_{\nu\lambda} H \gamma^{\mu 5} \gamma_\sigma\gamma_5 [\mathbb{A}^\sigma, P^{-}_{\lambda\nu}]\right)
      \biggr] \nonumber \\*&&{}+
      \cdots \Biggr\} \ .\eqn{L3A2}
\end{eqnarray}
The expressions inside of each square bracket are constructed by adding a derivative-containing
 factor in the same way to each of the single-trace or the double-trace terms of \eq{L2A2}, so the 
 ellipses  in the square brackets may easily be filled in if desired. On the other hand, the final
  ellipsis in \eq{L3A2} represents entirely new terms in which the operators breaking the 
  heavy-quark spin symmetry 
are contracted with $\mathbb{A}^\mu$ or $D^\mu$. An example is
\begin{equation}
  \Tr\left( \overline{H} \gamma_{5\mu} \xi_{\nu\lambda} H \{\mathbb{A}^\mu, P^{+}_{\lambda\nu}\}\right)\ .
\end{equation}

Similarly, for $\cL^{B2}_{3,a^2}$ we have:
\begin{eqnarray}\label{eq:L3B2}
\cL^{B2}_{3,a^2} & = &  a^2 \sum_\mu\Biggl\{ 
      \biggl[ i c^{B2}_{1,1} \Tr\left(\overline{H} \gamma_{\nu\mu} \xi_{\mu } H v\negcdot \leftvec D\, \gamma^{\mu\nu} P^{+}_{\mu}
		  - v\negcdot \rightvec D \, \overline{H} \gamma_{\nu\mu} \xi_{\mu } H \gamma^{\mu\nu} P^{+}_{\mu}\right) + \cdots 
      \biggr] \nonumber \\*&&{}+
      \biggl[ i c^{B2}_{2,1} \Tr\left(\overline{H} \gamma_{\nu\mu} \xi_{\mu } H v\negcdot \leftvec D\, \gamma^{\mu\nu} 
		  - v\negcdot \rightvec D \, \overline{H} \gamma_{\nu\mu} \xi_{\mu } H \gamma^{\mu\nu} \right) \Tr\left( P^{+}_{\mu}\right) + \cdots
      \biggr] \nonumber \\*&&{}+
      \biggl[ c^{B2}_{3,1} \Tr\left( \overline{H} \gamma_{\nu\mu} \xi_{\mu } H 
		  \gamma^{\mu\nu} \gamma_\sigma\gamma_5\{\mathbb{A}^\sigma, P^{+}_{\mu}\}\right) + \cdots
      \biggr] \nonumber \\*&&{}+
      \biggl[ c^{B2}_{4,1} \Tr\left( \overline{H} \gamma_{\nu\mu} \xi_{\mu } H 
		  \gamma^{\mu\nu} \gamma_\sigma\gamma_5\mathbb{A}^\sigma \right) \Tr\left( P^{+}_{\mu}\right) +  \cdots
      \biggr] \nonumber \\*&&{}+
      \biggl[ c^{B2}_{5,1} \Tr\left( \overline{H} \gamma_{\nu\mu} \xi_{\mu } H 
		  \gamma^{\mu\nu} \gamma_\sigma\gamma_5 \right) \Tr\left( \mathbb{A}^\sigma  P^{+}_{5}\right) + \cdots
      \biggr] \nonumber \\*&&{}+
      \biggl[ c^{A2}_{6,1} \Tr\left( \overline{H} \gamma_{\nu\mu} \xi_{\mu } H
		  \gamma^{\mu\nu} \gamma_\sigma\gamma_5 [\mathbb{A}^\sigma, P^{-}_{5}]\right)  +  \cdots 
      \biggr] \nonumber \\*&&{}+ 
	      \cdots \Biggr\}\  .
\end{eqnarray}

The case of the type-A contributions to the current, $j^{\mu,i \Xi}_{2,a^2,A2}$, is more 
straightforward, since we need only insert the heavy-quark and light-quark spurions, without
any additional derivatives, and Lorentz invariance is not broken.  Still, there are many terms, since
parity places no restrictions on the  low energy constants in the left-handed current, but merely 
relates them to those of the right-handed  current.  Further, many of the simplifying relations, 
\eqsthru{vslash-property1}{vslash-property6}, have no counterpart in the current, where there is
only a single heavy-meson field.
We therefore again only give some representative terms:
\begin{eqnarray}\label{eq:j2A2}
j^{\mu,i \Xi}_{2,a^2,A2} & = & a^2 \Biggl\{ 
       r^{A2}_{0,0}\, \trDt\left(\half T_\Xi\gamma^\mu (1-\gamma_5) H 
      \sigma^\dagger  \lambda^{(i)} \right ) +
       r^{A2}_{0,1}\, \trDt\left(\half T_\Xi\gamma^\mu (1-\gamma_5) \gamma_\nu H 
      \gamma^\nu \sigma^\dagger  \lambda^{(i)} \right ) \nonumber \\*&&{}+ 
      r^{A2}_{1,1}\, \trDt\left(\half T_\Xi\gamma^\mu (1-\gamma_5) \xi_{5}  H 
      P^{+}_{5}  \sigma^\dagger\lambda^{(i)} \right ) +
      r^{A2}_{1,2}\, \trDt\left(\half T_\Xi\gamma^\mu (1-\gamma_5) \xi_{\rho}  H 
      P^{+}_{\rho}  \sigma^\dagger\lambda^{(i)} \right )
      \nonumber \\*&&{}+     
      r^{A2}_{1,3}\, \trDt\left(\half T_\Xi\gamma^\mu (1-\gamma_5) \xi_{5\rho}  H 
      P^{+}_{\rho5}  \sigma^\dagger\lambda^{(i)} \right ) +
      r^{A2}_{1,4}\, \trDt\left(\half T_\Xi\gamma^\mu (1-\gamma_5) \xi_{\beta\rho}  H 
      P^{+}_{\rho\beta}  \sigma^\dagger\lambda^{(i)} \right )
      \nonumber \\*&&{}+
       r^{A2}_{1,5}\, \trDt\left(\half T_\Xi\gamma^\mu (1-\gamma_5) \gamma_\nu\xi_{5 } H \gamma^\nu
       P^{+}_5 \sigma^\dagger  \lambda^{(i)} \right )
       \nonumber \\*&&{}+ 
      r^{A2}_{1,6}\, \trDt\left(\half T_\Xi\gamma^\mu (1-\gamma_5) \gamma_{\nu\beta}\xi_{\rho}  H 
      \gamma^\beta v^\nu P^{+}_{\rho}  \sigma^\dagger\lambda^{(i)} \right ) 
     \nonumber \\*&&{}+
      r^{A2}_{1,7}\, \trDt\left(\half T_\Xi\gamma^\mu (1-\gamma_5) \gamma_{\nu\beta}\xi_{5\rho}  H 
      \gamma^\beta v^\nu P^{+}_{\rho5}  \sigma^\dagger\lambda^{(i)} \right ) 
     \nonumber \\*&&{}+
      r^{A2}_{1,8}\, \trDt\left(\half T_\Xi\gamma^\mu (1-\gamma_5) \gamma_\nu\xi_{\beta\rho}  H 
      \gamma^\nu P^{+}_{\rho\beta} \sigma^\dagger\lambda^{(i)} \right)\nonumber\\*&&{}+
      r^{A2}_{2,1}\, \trDt\left(\half T_\Xi\gamma^\mu (1-\gamma_5) \xi_{\rho}  H 
       \sigma^\dagger\lambda^{(i)} \right )\Tr\left( P^{+}_{\rho}\right) +\cdots
     \nonumber \\*&&{}+
       r^{A2}_{3,1}\, \trDt\left(\half T_\Xi\gamma^\mu (1-\gamma_5) \xi_{5}  H 
      P^{-}_{5}  \sigma^\dagger\lambda^{(i)} \right ) + \cdots
   \nonumber \\*&&{}+
      r^{A2}_{4,1}\, \trDt\left(\half T_\Xi\gamma^\mu (1-\gamma_5) \xi_{\rho}  H 
       \sigma^\dagger\lambda^{(i)} \right )\Tr\left( P^{-}_{\rho}\right) +\cdots \Biggr\}\  .
\end{eqnarray}
Here we have divided the terms into five sub-classes: 
terms with no $P^\pm$ factors,
single traces with $P^+$, double traces
with $P^+$,  single traces with $P^-$,  and double traces
with $P^-$. The terms with no factors of $P^\pm$ (coefficients $r^{A2}_{0,0}$ and $r^{A2}_{0,1}$)
are rather trivial and break no taste symmetries, 
although the second does break heavy-quark spin symmetry.
The ellipses in \eq{j2A2} may easily be filled based on the terms of the sub-class of single traces with $P^+$.
In deriving \eq{j2A2} we have used the fact that a factor of
$\gamma_5$ before or after the $H$ field has 
no (nontrivial) effect, due to the 
presence of the left projector, $(1-\gamma_5)$.  Thus, for example, terms generated by $[A\times S]$,
$[A\times P]$, and $[A\times T]$ are identical to those from $[V\times S]$,  $[V\times P]$, $[V\times T]$, respectively.  

As we will see more explicitly in the discussion of $j^{\mu,i \Xi}_{2,a^2,B2}$ that follows,
the Lorentz structures that follow $H$ in \eq{j2A2} are not fixed by the spurions, but can be any combination of the available four-vectors $\gamma^\alpha$ and $v^\lambda$ consistent with Lorentz
invariance.  For example, the factor $\gamma^\beta v^\nu$ following $H$ in the $r^{A2}_{1,6}$ term,
could also in principle be replaced by $\gamma^{\beta\nu}$.  However, such a term would vanish
due to the identity $\gamma^{\beta\nu} \gamma^\mu \gamma_{\nu\beta} = 0$.

Finally we turn to the type-B contributions to the current, $j^{\mu,i \Xi}_{2,a^2,B2}$.
The reasoning is very similar in principle to that for the type-B Lagrangian, but the presence
of an additional Lorentz index in the current increases the complexity, so we describe some
of the details.  Up to this point, we have not explicitly employed 
a formal spurion analysis for the current, but it now becomes necessary.
At the SET/HQET level, the left-handed current is
\begin{equation}
j^{\mu,i \Xi}=\bar q \big(\lambda^{(i)}\; \half  T^\Xi \gamma^\mu(1-\gamma_5)\big)Q =
\bar q \big(F(\mu)\otimes E\big)Q\ ,  \eqn{jSET}
\end{equation} 
where we have introduced a taste spurion $E$ and a spin spurion $F(\mu)$. They transform as
\begin{eqnarray}
E &\to & L E V^\dagger, \qquad  [\Rightarrow\ \sigma^\dagger E \to  \mathbb{U} \sigma^\dagger E V^\dagger] \ , \\
F(\mu) &\to & F(\mu) S^\dagger \ ,
\end{eqnarray}
and ultimately take the values
\begin{eqnarray}
E &= & \lambda^{(i)} \;\half T^\Xi \ , \\
F(\mu) &= & \gamma^\mu(1-\gamma_5) \ .
\end{eqnarray}

For an example, we again take the $[T_\mu \times A_\mu]$ type-B operator, and introduce spurions
for it as in \eq{T-mu*A-mu}, except we replace the index $\mu$ there with $\nu$ (and $\nu$ with
$\beta$)  so as not
to conflict with the index of the current. The terms we seek are trilinear in the spurions 
$F(\mu)\otimes E$, $B(\beta,\nu)\otimes C(\nu)$, and either $A_1(\nu)$ or $A_2(\nu)$.
Since parity does not constrain the terms in the current, we use just $A_1(\nu)$ in this
example.  Demanding heavy and light taste symmetry and heavy-quark spin symmetry, a possible
chiral operator has the form
\begin{equation}
\trDt\left\{\big(F(\mu)\otimes \sigma^\dagger E\big)\big(B(\beta,\nu)\otimes C(\nu)\big) H
                  \Gamma \sigma^\dagger A_1(\nu)  \sigma^\dagger  \right\}\ , \eqn{jBi}
\end{equation}
where $\mu$, $\nu$, and $\beta$ are fixed, and $\Gamma$ is some combination of components
of $\gamma$ matrices and of $v$, to be determined. 

After replacing the spin spurions $F(\mu)$
and $B(\beta,\nu)$ with their values, and reintroducing the sum over $\beta$, the current
becomes the $\;{}^\mu$ component of a Lorentz vector, and the 
4-quark operator becomes the $\;{}_\nu {}^\nu$ component of a symmetric two-index 
tensor.  As before, we may take the latter to be traceless.
At the SET level, call the three-index
tensor coming from the product of the 
two representations $X$.  As worked out in 
Appendix \ref{sec:tensor}, the element $X^{\mu\phantom{\nu}\nu}_{\phantom{\mu}\nu}$ is a linear
combination of elements of three irreducible
representations: a completely symmetric traceless three-index tensor ($S$), a three-index 
tensor with mixed
symmetry ($A$), and a vector ($W$). 
From Lorentz symmetry alone, the
chiral operators for each of these three representations 
could have independent LECs. Fixing the
spin spurions in \eq{jBi}, however, tells us that 
the corresponding chiral operator is required by spin symmetry to have the form 
\begin{equation}
\tilde X^{\mu\phantom{\nu}\nu}_{\phantom{\mu}\nu} =\trDt\left\{\sigma^\dagger E  \gamma^\mu(1-\gamma_5) \gamma_{\beta\nu} C(\nu) H\,
                  \Gamma^{\nu\beta} \sigma^\dagger A_1(\nu)  \sigma^\dagger  \right\}\ , \eqn{jBii}
\end{equation}
with an implicit sum over $\beta$, but not over $\nu$.
Given a choice for $\Gamma^{\nu\beta}$ (for example, $v^\nu v^\beta$), the corresponding elements of the individual representations at the chiral level,
$\tilde S^{\mu\phantom{\nu}\nu}_{\phantom{\mu}\nu}$, $\tilde A^{\mu\phantom{\nu}\nu}_{\phantom{\mu}\nu}$,
and $\tilde W^\mu$, which are formed by permuting indices and taking traces of $\tilde X$, will not in general
have the form of \eq{jBii} unless the properties of
$H$ and the Dirac trace conspire to allow them to be rewritten in that form.  We have checked that,
for the four possible choices for $\Gamma^{\nu\beta}$ ($\gamma^\nu\gamma^\beta$, $\gamma^\nu v^\beta$,
$v^\nu\gamma^\beta$, and $v^\nu v^\beta$), the generic situation obtains.%
\footnote{Note that the trivial choice  $\Gamma^{\nu\beta}=\delta^{\nu\beta}$ vanishes after the trace on
$\nu = \beta$ is subtracted, so only a type-A chiral operator can be formed in that way.}
Thus  the relative
normalization of the LECs of the individual representations are fixed to be the
same as in \eq{Xred}, and $\tilde X^{\mu\phantom{\nu}\nu}_{\phantom{\mu}\nu}$ is the only
possible chiral operator.  Setting the remaining spurions to their fixed values, and restoring the sum
over $\nu$, then gives the final chiral operators.  For the choice 
$\Gamma^{\nu\beta}=\gamma^\nu\gamma^\beta$, we find the operators
\begin{equation}
\trDt\left\{\half T_\Xi\gamma^\mu (1-\gamma_5) \gamma_{\nu\beta} \xi_{5\nu} H
		  \gamma^{\beta\nu} P^{\pm}_{\nu5} \sigma^\dagger  \lambda^{(i)}\right\}\ , \eqn{jBiii}
\end{equation}
where $P^{\pm}_{\nu5}$ arises from the sum and difference of \eq{jBii} with the corresponding
operator after the replacement $A_1(\nu) \to A_2(\nu)$. 

Following this procedure for other heavy-light terms in the SET, we then have 
\begin{eqnarray}\label{eq:j2B2}
j^{\mu,i \Xi}_{2,a^2,B2} & = &  a^2 \sum_\nu \Biggl\{ 
     r^{B2}_{1,1}\, \trDt\left(\half T_\Xi\gamma^\mu (1-\gamma_5) \gamma_{\nu\beta} \xi_{\nu} H
		  \gamma^{\beta\nu} P^{+}_{\nu} \sigma^\dagger  \lambda^{(i)}\right ) 
      \nonumber \\*&&{}+ 
      r^{B2}_{1,2}\, \trDt\left(\half T_\Xi\gamma^\mu (1-\gamma_5) \gamma_{\nu\beta} \xi_{5\nu} H
		  \gamma^{\beta\nu} P^{+}_{\nu5} \sigma^\dagger  \lambda^{(i)}\right )
	 \nonumber \\*&&{}+
	r^{A2}_{1,3}\, \trDt\left(\half T_\Xi\gamma^\mu (1-\gamma_5) \gamma_\nu\xi_{\nu\rho}  H 
      \gamma^\nu P^{+}_{\rho\nu} \sigma^\dagger\lambda^{(i)} \right)
       \nonumber \\*&&{}+ 
       r^{A2}_{1,4}\, \trDt\left(\half T_\Xi\gamma^\mu (1-\gamma_5) \gamma_\nu\xi_{\nu\rho}  H 
       v^\nu P^{+}_{\rho\nu} \sigma^\dagger\lambda^{(i)} \right) + \cdots
       \nonumber \\*&&{}+ 
      r^{B2}_{2,1}\, \trDt\left(\half T_\Xi\gamma^\mu (1-\gamma_5) \gamma_{\nu\beta} \xi_{\nu} H 
		  \gamma^{\beta\nu} \sigma^\dagger\lambda^{(i)}\right) \Tr\left( P^{+}_\nu \right )+ \cdots
       \nonumber \\*&&{}+ 
      r^{B2}_{3,1}\, \trDt\left(\half T_\Xi\gamma^\mu (1-\gamma_5) \gamma_{\nu\beta} \xi_{\nu} H
		  \gamma^{\beta\nu} P^{-}_\nu \sigma^\dagger  \lambda^{(i)}\right )+ \cdots
       \nonumber \\*&&{}+                
    r^{B2}_{4,1}\, \trDt\left(\half T_\Xi\gamma^\mu (1-\gamma_5) \gamma_{\nu\beta} \xi_{\nu} H 
		  \gamma^{\beta\nu} \sigma^\dagger\lambda^{(i)} \right)\Tr\left( P^{-}_\nu \right )+ \cdots\Biggr\}
     \  , 
\end{eqnarray}
where again we have not written the complete set of contributions, but only some representative terms.

One can use a similar spurion analysis to check the type-B contributions to the current coming from the
light-light four-quark operators, \eq{j2B1}.  In that case, the only Dirac matrix coming
 before the $H$ field is the $\gamma^\mu(1-\gamma_5)$
spin spurion from the current, and the  matrix corresponding to 
$\Gamma^{\nu\beta}$ after $H$ is simply the $\;{}_\nu {}^\nu$ component of a two-index symmetric, traceless
tensor.  The choices $v_\nu v^\nu$ and $v_\nu\gamma^\nu$ for this matrix ($\gamma_\nu\gamma^\nu$ is
clearly trivial) give the terms in  \eq{j2B1}.   The incorrect additional terms listed in \rcite{Aubin:StagHL2006}
came from ignoring the consequences of heavy-quark spin symmetry, and using Lorentz-symmetry
considerations only.

This completes the discussion of the effects of light-heavy terms in the SET, \eq{SET_lh}.   There are still the
heavy-heavy terms, \eq{SET_hh} to consider.  However, it is now easy to see that the heavy-heavy terms
do not produce any new nontrivial chiral
operators in the Lagrangian or current.  These 4-quark operators contain two heavy-quark spurions, and no light-quark spurions.
Since the heavy-quark spurions transform on both sides with heavy-quark spin matrices and heavy-quark taste matrices, they both must be placed between the $\overline{H}$ and $H$ fields in the Lagrangian.  One then just gets the product of the two spurions, which is proportional to the identity.  So the heavy-heavy 4-quark operators in the SET
lead simply to trivial chiral Lagrangian operators, which are already present as the first operators in \eqs{L2A2}{L3A2}.
For the same reason, they lead to a trivial current operator, $a^2 \trDt\left(\frac{1}{2}T_\Xi\gamma^\mu (1-\gamma_5) H \sigma^\dagger  \lambda^{(i)} \right )$, which does not break any symmetries and just adds a constant term proportional to $a^2$ to any LO matrix element.

\section{Taste splittings of heavy-light meson masses}
\label{sec:Mass}

In this section, we calculate the mass splitting between heavy-light mesons of different tastes in terms of
the low energy constants in the chiral Lagrangian.  With reasonable assumptions about which operators give dominant effects, we are able to explain the observed pattern of taste splittings.

We first show that the one-loop diagrams give taste-invariant masses 
to the heavy-light mesons, even though the diagrams contain pion propagators, which  
break taste symmetry. 
Taste-independence of one-loop chiral logs follows from the exact $SU(4)$ taste symmetry of the heavy quark at LO in the chiral theory, 
as well as the shift symmetry of the staggered action \cite{SHIFT}.  The latter
can be represented at the SET and chiral levels as an exact, discrete taste symmetry that 
acts jointly on both heavy and light quarks \cite{BGS08}.
This symmetry is
\begin{eqnarray}
  q_i \to (I\otimes \xi_\nu) q_i \,,&\quad&
  \bar q_i \to \bar q_i (I \otimes \xi_\nu) \ ,
  \label{eq:disc_shift1} \\
  Q \to (I\otimes \xi_\nu) Q \,,&\quad&
  \bar Q \to \bar Q (I \otimes \xi_\nu) \ ,
  \label{eq:disc_shift2}
\end{eqnarray}
at the level of the Symanzik action, and
\begin{eqnarray}\label{eq:shift_chiral}
  \Sigma & \to & \xi^{(n)}_\nu \Sigma \xi^{(n)}_\nu\ , \nonumber\\
  \sigma & \to & \xi^{(n)}_\nu \sigma \xi^{(n)}_\nu\ , \nonumber\\
  H      & \to & \xi_\nu H \xi^{(n)}_\nu \ ,\nonumber\\
  \overline{H}    & \to & \xi^{(n)}_\nu \overline{H} \xi_\nu \ ,
\end{eqnarray}
at the chiral level.
Note that the symmetry is diagonal in flavor;
the transformation acts only on the taste
indices and affects all light quark flavors, as well as the heavy quark, identically.

Using the $SU(4)$ heavy-quark taste symmetry of the LO Lagrangian,
one can undo the action of the discrete taste symmetry on the heavy quark. 
Taking $V=\xi_\nu$ in \eq{Htaste}, we have the following symmetry of the LO Lagrangian:
\begin{eqnarray}\label{eq:shift_chiral}
  \Sigma & \to & \xi^{(n)}_\nu \Sigma \xi^{(n)}_\nu\ , \nonumber\\
  \sigma & \to & \xi^{(n)}_\nu \sigma \xi^{(n)}_\nu\ , \nonumber\\
  H      & \to & H \xi^{(n)}_\nu \ ,\nonumber\\
  \overline{H}    & \to & \xi^{(n)}_\nu \overline{H} \ ,
\end{eqnarray}
We call this symmetry {\it light-quark discrete taste symmetry}\/.
In applying it, it is convenient to think of $H$ in the way described above
\eq{L1}, as a light flavor vector (index $i$) with components that are
$4\times4$ taste matrices
\begin{equation}
H_i^{\alpha\beta} = \sum_{\Xi=1}^{16}\half T_\Xi^{\alpha\beta}\,  H_{i\Xi\;}\ .
\end{equation}
Here $\alpha$ and $\beta$ are the heavy and light quark tastes, respectively. 

We can now show that the heavy-light meson propagator is taste invariant if the
$SU(4)$ heavy-quark taste symmetry is exact.  This implies that the one-loop
diagrams for the propagator are taste invariant, since they use LO propagators
and vertices. Consider the propagator
\begin{equation}\eqn{H-prop}
\langle 0 \vert H_i^{\alpha\beta}(x) \overline{H}_j^{\beta'\alpha'}(y) \vert0 \rangle\equiv
\delta_{ij}K^{\alpha\alpha'}(\beta,\beta',x,y,i)\;,
\end{equation}
where we have used flavor conservation. Then the heavy taste symmetry implies
\begin{equation}
K^{\alpha\alpha'} = (VKV^\dagger)^{\alpha\alpha'} 
\end{equation}
for any $SU(4)$ taste transformation $V$. Thus $K$ is proportional to the identity, 
which gives
\begin{equation}
\langle 0 \vert H_i^{\alpha\beta}(x) \overline{H}_j^{\beta'\alpha'}(y) \vert0 \rangle\equiv
\delta_{ij}\delta^{\alpha\alpha'}G^{\beta'\beta}(x,y,i)=
\delta_{ij}\delta^{\alpha\alpha'}\sum_{\Xi=1}^{16}\half  T_\Xi^{\beta'\beta}\,
g_\Xi(x,y,i)\ ,
\end{equation}
where we have defined (equivalent) new functions $G^{\beta'\beta}$ and  $g_\Xi$. Light-quark discrete taste symmetry, \eq{shift_chiral}, implies
\begin{equation}
\sum_{\Xi=1}^{16} T_\Xi
\,
g_\Xi(x,y,i) = \sum_{\Xi=1}^{16} \xi_\nu T_\Xi
\xi_\nu\,
g_\Xi(x,y,i)\eqn{shift-consequence}\;.
\end{equation}
Each $T_\Xi$ has a unique signature of four signs determined
by whether $ \xi_\nu T_\Xi \xi_\nu$ is $+T_\Xi$ or $-T_\Xi$, for $\nu=1,\cdots,4$.
Clearly only $T_\Xi=I$ has signature $(+,+,+,+)$.
One may then conclude from \eq{shift-consequence} that
$g_\Xi=0$ for $\Xi\not=I$ and
\begin{equation}
\langle 0 \vert H_i^{\alpha\beta}(x) \overline{H}_j^{\beta'\alpha'}(y) \vert0 \rangle=
\half\delta_{ij}\delta^{\alpha\alpha'}\,
\delta^{\beta\beta'}g_I(x,y,i)\ .
\end{equation}
Multiplying with $\half T_\Xi^{\beta\alpha}$ and $\half T_{\Xi'}^{\alpha'\beta'}$ and summing
repeated indices gives the final form
\begin{equation}
\langle 0 \vert H_{i\Xi}(x) \overline{H}_{j\Xi'}(y) \vert0 \rangle=
\half\delta_{ij}\delta_{\Xi\Xi'}\,
g_I(x,y,i)\ .
\end{equation}
Thus the one-loop heavy-light meson propagator is taste invariant, so the masses
(as well as the wave function renormalization) at one-loop are invariant.
This means that all taste-violations
in the heavy-light masses at NLO come from the NLO terms in the \aschpt\ Lagrangian, treated at tree level,
and may be analyzed straightforwardly.

From now on we refer to the heavy-light  pseudoscalar meson as a $D$ (not $B$) meson, 
because the lattice data from MILC that we show later is for $D$ mesons.  
To determine the taste splittings in the meson masses, we need only
consider the taste-violating NLO Lagrangian terms $\cL^{A1}_{2,a^2}$,  $\cL^{B1}_{2,a^2}$,  $\cL^{A2}_{2,a^2}$
and $\cL^{B2}_{2,a^2}$.  Taste-violating terms in $\cL_3$ lead only to wave-function renormalization, since the LO pole in the propagator is at residual 
momentum $k=0$, and these terms either have an addition factor of $k$ or at least 
one pion field.   Further, one easily sees that  $\cL^{A1}_{2,a^2}$ and $\cL^{B1}_{2,a^2}$, \eqs{L2A1}{L2B1}, produce no
taste splittings of $D$ mesons because their taste-noninvariant factors, $\cO^{A1,+}_k$ and
$\cO^{B1,+}_{\mu,k}$ [\eqs{OA1}{OB1}], either vanish or go to the identity matrix when there are no pion 
fields at tree level.  Thus taste splittings of $D$ meson masses at NLO (\ie $\cO(a^2)$) come only from the terms that
break heavy-quark taste and spin symmetry, namely $\cL^{A2}_{2,a^2}$
and $\cL^{B2}_{2,a^2}$.
From \eqs{L2A2}{L2B2}, we can then easily find all the terms that
contribute to taste splittings of the $D$ masses at $\cO(a^2)$:
\begin{eqnarray}
\delta \cL_{m_Q} & = & 
  a^2 \Biggl\{K^{A2}_{1,1} \Tr\left(\overline{H} \xi_{5   } H \xi_{    5} \right) +
	      K^{A2}_{1,2} \Tr\left(\overline{H} \xi_{ \mu} H \xi_{\mu  } \right) \nonumber \\*&&{}+
	      K^{A2}_{1,3} \Tr\left(\overline{H} \xi_{5\mu} H \xi_{\mu 5} \right) +
	      K^{A2}_{1,4} \Tr\left(\overline{H} \xi_{\mu\nu} H \xi_{\nu\mu} \right) \nonumber \\*&&{}+
	      K^{A2}_{1,6}\Tr\left(\overline{H} \gamma_{5\mu} \xi_{5 } H  \gamma^{\mu 5} \xi_{ 5} \right)  +
	      K^{A2}_{1,7} \Tr\left(\overline{H} \gamma_{\mu\nu} \xi_{\lambda } H  \gamma^{\nu\mu} \xi_{\lambda} \right) \nonumber \\*&&{}+
	      K^{A2}_{1,8} \Tr\left(\overline{H} \gamma_{\mu\nu} \xi_{5\lambda } H  \gamma^{\nu\mu} \xi_{\lambda 5} \right) +
	      K^{A2}_{1,9} \Tr\left(\overline{H} \gamma_{5\mu} \xi_{\nu\lambda} H  \gamma^{\mu 5} \xi_{\lambda\nu} \right) \Biggr\} \nonumber \\*&&{}+
  a^2 \sum_\mu \Biggl\{ K^{B2}_{1,1} \Tr\left(\overline{H} \gamma_{\nu\mu} \xi_{\mu }   H  \gamma^{\mu\nu} \xi_{\mu }  \right) +
	      K^{B2}_{1,2} \Tr\left(\overline{H} \gamma_{\nu\mu} \xi_{5\mu } H  \gamma^{\mu\nu} \xi_{\mu 5} \right)  \nonumber \\*&&{}+
	      K^{B2}_{1,3} v^\mu v_\mu \Tr\left(\overline{H} \xi_{\nu\mu} H \xi_{\mu\nu} \right) +
	      K^{B2}_{1,4} \Tr\left(\overline{H} \gamma_{5\mu} \xi_{\nu\mu } H  \gamma^{\mu 5} \xi_{\mu\nu} \right) \Biggr\} \ ,\label{eq:delta-L_M}
\end{eqnarray}
where we have set the pion fields $\sigma$ and $\sigma^\dagger$ to the identity.
The sum of the mass contributions from these terms has the form 
\begin{eqnarray}
\delta \cL_{m_Q} =  - \sum_{\Xi} D_{\Xi}^\dagger D_\Xi \triangle_{m_Q}(T_\Xi) + \cdots\;, 
\end{eqnarray}
where $\triangle_{m_Q}(T_\Xi)$ is the mass shift of the $D$ meson with taste $\Xi$, and $\cdots$ represents $D^*$ mass terms, which we are not interested in here.

For a static $D$ meson, where $v_i = 0$, the  corrections on the $D$ masses from $ \delta \cL_{m_Q}$ are:
\begin{eqnarray}
\triangle_{m_Q}(\xi_5)    & = & 2 a^2\Big\{(K^{A2}_{1,1}-3K^{A2}_{1,6})-4(K^{A2}_{1,2}+6K^{A2}_{1,7})-4(K^{A2}_{1,3}+6K^{A2}_{1,8})  \nonumber \\*&&{}
	  +12(K^{A2}_{1,4}-3K^{A2}_{1,9})      -6K^{B2}_{1,1}-6K^{B2}_{1,2}+3K^{B2}_{1,3}-9K^{B2}_{1,4}\Big\} \\
\triangle_{m_Q}(\xi_{05}) & = & 2 a^2\Big\{-(K^{A2}_{1,1}-3K^{A2}_{1,6})+2(K^{A2}_{1,2}+6K^{A2}_{1,7})-2(K^{A2}_{1,3}+6K^{A2}_{1,8}) \nonumber \\*&&{}
	  +6K^{B2}_{1,1}-6K^{B2}_{1,2}-3K^{B2}_{1,3}-3K^{B2}_{1,4}\Big\} \\ 
\triangle_{m_Q}(\xi_{i5}) & = & 2 a^2\Big\{-(K^{A2}_{1,1}-3K^{A2}_{1,6})+2(K^{A2}_{1,2}+6K^{A2}_{1,7})-2(K^{A2}_{1,3}+6K^{A2}_{1,8}) \nonumber \\*&&{}
	  +2K^{B2}_{1,1}-2K^{B2}_{1,2}+K^{B2}_{1,3}+K^{B2}_{1,4}\Big\} \\
\triangle_{m_Q}(\xi_{ij}) & = & 2 a^2\Big\{(K^{A2}_{1,1}-3K^{A2}_{1,6}) -4(K^{A2}_{1,4}-3K^{A2}_{1,9})  -2K^{B2}_{1,1}-2K^{B2}_{1,2} \nonumber \\*&&{}
	  -K^{B2}_{1,3}+3K^{B2}_{1,4}\Big\} \\
\triangle_{m_Q}(\xi_{i0}) & = & 2 a^2\Big\{(K^{A2}_{1,1}-3K^{A2}_{1,6}) -4(K^{A2}_{1,4}-3K^{A2}_{1,9})  +2K^{B2}_{1,1}+2K^{B2}_{1,2} \nonumber \\*&&{}
	  -K^{B2}_{1,3}+3K^{B2}_{1,4}\Big\} \\
\triangle_{m_Q}(\xi_i)    & = & 2 a^2\Big\{-(K^{A2}_{1,1}-3K^{A2}_{1,6})-2(K^{A2}_{1,2}+6K^{A2}_{1,7})+2(K^{A2}_{1,3}+6K^{A2}_{1,8}) \nonumber \\*&&{}
	  -2K^{B2}_{1,1}+2K^{B2}_{1,2}+K^{B2}_{1,3}+K^{B2}_{1,4}\Big\} \\
\triangle_{m_Q}(\xi_0)    & = & 2 a^2\Big\{-(K^{A2}_{1,1}-3K^{A2}_{1,6})-2(K^{A2}_{1,2}+6K^{A2}_{1,7})+2(K^{A2}_{1,3}+6K^{A2}_{1,8}) \nonumber \\*&&{}
	  -6K^{B2}_{1,1}+6K^{B2}_{1,2}-3K^{B2}_{1,3}-3K^{B2}_{1,4}\Big\} \\
\triangle_{m_Q}(I)        & = & 2 a^2\Big\{(K^{A2}_{1,1}-3K^{A2}_{1,6})+4(K^{A2}_{1,2}+6K^{A2}_{1,7})+4(K^{A2}_{1,3}+6K^{A2}_{1,8})  \nonumber \\*&&{}
	  +12(K^{A2}_{1,4}-3K^{A2}_{1,9})      +6K^{B2}_{1,1}+6K^{B2}_{1,2}+3K^{B2}_{1,3}-9K^{B2}_{1,4}\Big\}  
\end{eqnarray}
The results are summarized in {Tables \ref{tab:TasteSplittingA} and \ref{tab:TasteSplittingB}}, which help us see the patterns of taste splittings. 
\begin{table}
\caption{Taste splittings due to type-A operators}
\label{tab:TasteSplittingA}
\begin{tabular}{|c||*{5}{ >{\centering\arraybackslash}p{1cm}|} } \hline
{$\triangle_{m_Q}(.)$} & $\xi_5$ & $\xi_{\mu5}$  & $\xi_{\mu\nu}$ & $\xi_\mu$ &  $I$ \\ \hline \hline
$2 a^2 (K^{A2}_{1,1}-3K^{A2}_{1,6})$  &  +1  & -1 & +1 & -1 & +1 \\ \hline   
$2 a^2 (K^{A2}_{1,2}+6K^{A2}_{1,7})$  &  -4  & +2 &  0 & -2 & +4 \\ \hline 
$2 a^2 (K^{A2}_{1,3}+6K^{A2}_{1,8})$  &  -4  & -2 &  0 & +2 & +4 \\ \hline 
$2 a^2 (K^{A2}_{1,4}-3K^{A2}_{1,9})$  &  +12  &  0 & -4 &  0 & +12 \\ \hline 
\end{tabular}
\end{table}
\vspace{.2in}

\begin{table}
\caption{Taste splittings due to type-B operators}
\label{tab:TasteSplittingB}
\begin{tabular}{|c||>{\centering\arraybackslash}p{1cm}|
			*{3}{*{2}{ >{\centering\arraybackslash}p{1cm}}|}>{\centering\arraybackslash}p{1cm}|} \hline
$\triangle_{m_Q}(.)$  & $\xi_5$ & $\xi_{05}$ & $\xi_{i5}$ & $\xi_{ij}$ & $\xi_{i0}$ & $\xi_i$ & $\xi_0$ & $I$ \\ \hline \hline
$2 a^2 K^{B2}_{1,1}$  &  -6  &  +6  &  +2  &  -2  & +2 & -2 & -6 & +6 \\ \hline   
$2 a^2 K^{B2}_{1,2}$  &  -6  &  -6  &  -2  &  -2  & +2 & +2 & +6 & +6 \\ \hline 
$2 a^2 K^{B2}_{1,3}$  &  +3  &  -3  &  +1  &  -1  & -1 & +1 & -3 & +3 \\ \hline 
$2 a^2 K^{B2}_{1,4}$  &  -9  &  -3  &  +1  &  +3  & +3 & +1 & -3 & -9 \\ \hline 
\end{tabular}
\end{table}

The type-A terms split the heavy-light masses into the five $SO(4)$ taste multiplets: P, A, T, V and S 
(pseudoscalar, axial-vector, tensor, vector and singlet tastes).
The type-B terms split these multiplets and give different masses 
to the time and spatial components, such as $\xi_0$ and $\xi_i$ for the vector
taste multiplet.  The staggered lattice symmetries guarantee that the eight multiplets shown in 
\tabref{TasteSplittingB} cannot be be broken further; for example, the three tastes $\xi_i$ must
remain degenerate.   On the other hand, 
it is straightforward to check that any pattern of splitting of the eight multiplets is possible, given arbitrary values of the parameters $K^{A2}_{1,n}$ and 
$K^{B2}_{1,m}$.  

Further progress in understanding the actual pattern of splittings determined in
simulations is therefore only possible with some assumptions about which of the corresponding
chiral operators are likely to give dominant contributions to the masses. Experience with the pion (light-light
pseudoscalar)
splittings is helpful in guiding these assumptions, so we first review what happens
in that case. The staggered pion masses at LO are
\begin{equation}
 m^2_{ab,\Xi} = \mu(m_a+m_b) + a^2\Delta_\Xi\ ,
\end{equation}
where $m_a$ and $m_b$ are light quark masses, $\mu$ is the low-energy constant
from \eq{Lpion}, and  $a^2\Delta_\Xi$
is the splitting of taste $\Xi$. 
The pions have $SO(4)$ taste symmetry; their masses form five multiplets with tastes P, A, T, V and S.
Simulations with the asqtad and HISQ actions 
give approximately equal splittings of squared masses between 
the P, A, T, V and S tastes (and with that ordering, from lowest to highest)  
\cite{Bazavov:2009bb,Bazavov:2010ru,Bazavov:2012xda}.
These equal splittings imply that the dominant chiral operator 
contributing to taste splittings of pion masses is the operator multiplied by $C_4$ in $-a^2V_\Sigma$,
\eq{V}, namely
\begin{equation}\label{eq:dominant_pionmass_splitting}
    a^2 \left[\Tr(\xi^{(n)}_{\nu 5}\Sigma \xi^{(n)}_{5\nu}\Sigma) + h.c.\right] \ .
\end{equation}
This operator is generated by the four-quark operators
$[S\times A]^{ll}$, $[P\times A]^{ll}$ and $[T\times A]^{ll}$ in the SET, \eq{SET_ll}.
Note that, for the pions, only type-A operators are relevant at LO, because type-B 
operators have no chiral representatives to this order.
The non-trivial space-time structure in the type-B case requires more at least two derivatives 
in the light-light chiral operators, making their representatives NLO in the chiral expansion 
\cite{LEE_SHARPE}.

We now carry over this experience to the heavy-light case. We have assumed above that
the lattice is sufficiently fine, or the charmed quark is sufficiently improved, that it may
treated as a ``continuum-like,'' and corrections of order $(am_Q)^2$ may be neglected. This
means that the contributions of the heavy quark to the SET are identical to those of a light
quark. In particular, the same four-quark operators that dominated for light quarks, namely
$[S\times A]$, $[P\times A]$ and $[T\times A]$, are expected to
 be the dominant type-A operators in the heavy-light case.
Taste splittings of heavy-light meson masses can come only from the ``heavy-light'' versions
of these operators. From \eqsthree{SxA}{PxA}{TxA}, these operators give rise to chiral representatives
with coefficients $K^{A2}_{1,3}$ and $K^{A2}_{1,8}$ in \eq{delta-L_M}.
From \tabref{TasteSplittingA}, we then deduce 
the same equal-spacing 
pattern for heavy-light $SO(4)$ representations that is familiar from the 
pions.  For type-B operators, one may guess that the 
$[T_\mu\times A_\mu]$ SET operator would be dominant, since
it is the only type-B operator that has the same spin and taste as one of the dominant
type-A operators. From \eq{TmuxAmu}, this four-quark operator gives rise to the chiral representative
with coefficient $K^{B2}_{1,2}$ in  \eq{delta-L_M}.  Referring to the second line of
Table~\ref{tab:TasteSplittingB}, we see that this operator produces equal splitting within the 
A, T, and V $SO(4)$ multiplets.  Further, the multiplicity-weighted average splitting between $SO(4)$
multiplets for this type-B operator
is the same as for the dominant type-A operators (equal splitting with the order P, A, T, V, S), so
this operator does not spoil that overall $SO(4)$ pattern, but only produces splittings within multiplets.

\begin{figure}[thbp]
\begin{center}
\includegraphics[width=4in]{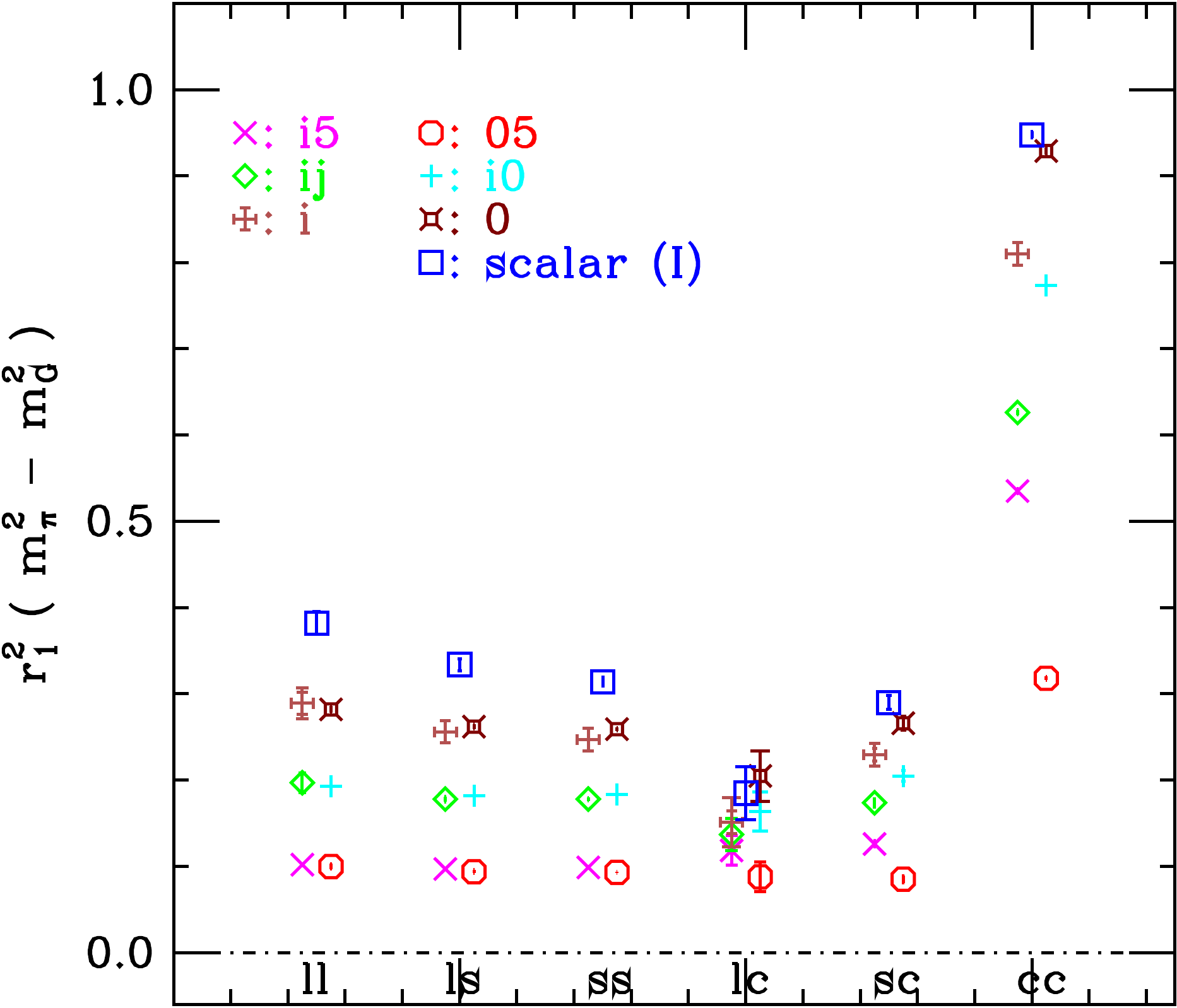}
\end{center}
\caption{Meson mass splitting for the MILC HISQ ensemble at $a\approx0.15\;$fm and $m_l=0.2m_s$ \cite{Bazavov:2012xda}.
Squared mass splitting between pions of different tastes and the Goldstone pion in units of
$r_1$ are shown.  The types of quarks in the mesons are shown on the abscissa: l, s, and c stand for
light (u,d), strange, and charm quarks, respectively.}
\label{fig:MediumCoarseMILCensemble}
\end{figure}

The patterns of splitting expected from the discussion in the previous paragraph 
are qualitatively present in the MILC data, shown
in \figref{MediumCoarseMILCensemble}.  Note in particular the ``sc'' case, which gives heavy-light meson
splittings with small enough
errors that the pattern of $SO(4)$ breaking
is clear.  It is non-trivial that the time component of taste is higher than the space
components in two cases ($\xi_0$ \vs $\xi_i$ and $\xi_{i0}$  \vs $\xi_{ij}$) but not
in the third case ($\xi_{05}$  \vs $\xi_{i5}$), just as in the second line
of Table~\ref{tab:TasteSplittingB}.  Further, the figure shows roughly equal splittings within $SO(4)$
multiplets, as well as between (the center of gravity of) $SO(4)$ multiplets. 
Although the chiral theory is
not applicable to the ``cc'' case, it is interesting to see that the structure
that would correspond to the dominant type-B operator 
gets particularly strong there, with near degeneracies of between members of different
$SO(4)$ multiplets, in particular $\xi_0$ and $I$, or $\xi_{i0}$ and $\xi_i$.

\section{Decay constants of the $D$ meson at NLO}
\label{sec:fD}

In this section, we calculate the decay constant of the $D$ meson at one loop in
\aschpt. We can express  the decay constant at this order as
\begin{equation}\label{eq:fD}
        f_{D_{x\Xi}}\sqrt{M_{D_{x\Xi}}} = \kappa \left( 1 + \frac{1}{16\pi^2 f^2}\;
        \delta\! f_{D_{x}} + {\rm analytic\ terms} \right) \ ,
\end{equation}
where $x$ labels the light valence flavor in the meson, $\Xi$ labels the meson taste,
$\kappa$ is the LO low-energy constant in the current, \eq{LOcurrent},
and $\delta\! f_{D_{x}} $ denotes the sum of the chiral logarithm terms, coming from the one-loop
diagrams.  We will allow for the 
possibility of partial quenching, so the valence quark mass $m_x$ may be different from any 
of the sea-quark masses.
The analytic terms arise from tree-level contributions from the NLO Lagrangian and
current and will include taste symmetry violations, 
due to the taste-violating terms  $\cL^{A2}_{3,a^2}\ $, $\cL^{B2}_{3,a^2}\ $, $j^{\mu,i \Xi}_{2,a^2,A2}\ $ 
and $j^{\mu,i \Xi}_{2,a^2,B2}\ $.

By following the same approach as we used to show the one-loop contribution to the
heavy-light meson propagator is taste independent, it is straightforward to show that the one-loop term $\delta\! f_{D_{x}}$  is independent of taste of the meson.  We simply replace the field $H_i^{\alpha\beta}$ in \eq{H-prop} with the leading order current
\begin{equation}\label{eq:j-prop}
  j^{\mu,i,\alpha\beta}_{\rm LO}= \sum_\Xi \half T_\Xi^{\alpha\beta} j^{\mu,i \Xi}_{\rm LO} =  \frac{\kappa}{2} \sum_\Xi \half T_\Xi^{\alpha\beta}\; 
  \trDt\bigl(\half T_\Xi\gamma^\mu\left(1-\gamma_5\right)H\sigma^\dagger\lambda^{(i)} \bigr)\ ,
\end{equation}
where we have used \eq{LOcurrent} for $j^{\mu,i \Xi}_{\rm LO}$. 
Note that $ j^{\mu,i,\alpha\beta}_{\rm LO}$ transforms under  heavy-quark taste 
symmetry and  light-quark discrete taste symmetry exactly as $H_i^{\alpha\beta}$
does. Identical manipulations
to those in \secref{Mass} thus show that the two-point function
of the current and the field
is taste-independent:
\begin{equation}\eqn{j-prop-final}
\langle 0 \vert  j^{\mu,i \Xi}_{\rm LO} (x) \overline{H}_{j\Xi'}(y) \vert0 \rangle=
\half\delta_{ij}\delta_{\Xi\Xi'}\,
h_I(x,y,i)\ ,
\end{equation}
where we have introduced a new function $h_I$.    
Up to an additional term 
coming from the one-loop wave function renormalization, $\delta\! f_{D_{x}}$ 
is proportional to the one-loop contribution to the two-point function in \eq{j-prop-final}.
Since we know from \secref{Mass} that the wave function contribution is taste-independent, we have proven the taste-independence of  $\delta\! f_{D_{x}}$.
Furthermore, it is now easy to see that $\delta\! f_{D_{x}}$ in our theory,
\aschpt, is identical to the corresponding contribution in \rhmschpt\ calculated
in  Ref.~\cite{Aubin:StagHL2006}.  The only difference between the LO Lagrangians
in the two theories is the extra taste degree of freedom of the heavy quark
in \aschpt. Since we
have seen that heavy-quark taste is conserved in the one-loop diagrams, the heavy taste
degree of freedom just flows through the diagram and has no effect on the result.
Note that virtual heavy quark
loops are forbidden in our theory since the residual energy is low; if they were allowed
the heavy-quark taste would lead to an extra counting factor in loops.

We thus take over the result from Ref. \cite{Aubin:StagHL2006} 
for $\delta\! f_{D_{x}}$ without change, except for trivial changes in notation.
The analytic terms, which come from the NLO Lagrangian, will be different
in the two theories, however.
The terms 
$\cL^{A2}_{3,a^2}\ $, $\cL^{B2}_{3,a^2}\ $, $j^{\mu,i \Xi}_{2,a^2,A2}\ $ 
and $j^{\mu,i \Xi}_{2,a^2,B2}\ $ give contributions that depend on the taste
of meson.

Following Ref.~\cite{Aubin:StagHL2006} for the one loop terms, we then get, for the
\opopo\ partially quenched case with all masses unequal:
\begin{eqnarray}\label{eq:1p1p1_pq_fD}
 \frac{ f_{D_{x\Xi}} \sqrt{ M_{D_{x\Xi}} } }{\kappa}
   &= & 1 + \frac{1}{16\pi^2f^2}
      \frac{1+3g_\pi^2}{2}
      \Biggl\{-\frac{1}{16}\sum_{\mathscr{S},\Xi'} \ell(m_{x\mathscr{S},\Xi'}^2)  \nonumber \\*&&{}-
	\frac{1}{3}\sum_{j\in \cM_I^{(3,x)}} 
	\frac{\partial}{\partial m^2_{X,I}}\left[ R^{[3,3]}_{j}(
	  \cM_I^{(3,x)};  \mu^{(3)}_I)\ell(m_{j}^2) \right] \nonumber \\*&&{}- 
	\biggl( a^2\delta'_V \sum_{j\in \cM_V^{(4,x)}}
	\frac{\partial}{\partial m^2_{X,V}}
	\left[ R^{[4,3]}_{j}( \cM_V^{(4,x)}; \mu^{(3)}_V )
	\ell(m_{j}^2) \right]
	    + [V\to A] \biggr)
      \Biggr\}  \nonumber \\*&&{}+
       c_s (m_u + m_d + m_s) + c_v m_x + c_{a,\Xi} a^2 \ ,
\end{eqnarray}
where x is the valence flavor, $\Xi$ is the valence taste, $\mathscr{S}$ 
runs over the three sea quarks $u$, $d$, and $s$, 
and $\Xi'$ runs over the 16 meson tastes. The chiral logarithm function $\ell$ and the residue functions
$R$ are defined by
\begin{eqnarray}
        \ell(m^2) &\equiv & m^2 \ln \frac{m^2}{\Lambda_\chi^2} , \label{eq:chiral_log_infinitev} \\
        R_j^{[n,k]}\left(\left\{m\right\}\!;\!\left\{\mu\right\}\right)
                  &\equiv & \frac{\prod_{i=1}^k (\mu^2_i- m^2_j)}
		   {\prod_{r\not=j} (m^2_r - m^2_j)}\ ,
\end{eqnarray}  
with the sets of masses in the residues given by
\begin{eqnarray}
  \mu^{(3)} & = & \{m^2_U,m^2_D,m^2_S\}\ ,\\*
  \cM^{(3,x)} & = & \{m_X^2,m_{\pi^0}^2, m_{\eta}^2\}\ ,\\*
  \cM^{(4,x)} & = & \{m_X^2,m_{\pi^0}^2, m_{\eta}^2, m_{\eta'}^2\}\ .
\end{eqnarray}
Here taste labels (\eg $I$ or $V$ for the masses) are implicit.
In \eq{1p1p1_pq_fD}, $c_{a,\Xi}$ is the only coefficient that depends on the taste of the heavy meson.
It can be written as a linear function of  constants appearing in
 $\cL^{A2}_{3,a^2}\ $, $\cL^{B2}_{3,a^2}\ $, $j^{\mu,i \Xi}_{2,a^2,A2}\ $ and $j^{\mu,i \Xi}_{2,a^2,B2}\ $.  
 It is straightforward to check that these terms are sufficient to
 break the taste symmetry down to the lattice symmetry.  Thus the
coefficients $c_{a,\Xi}$ are independent for the eight multiplets listed in
\tabref{TasteSplittingB}.

Now we include the effects of hyperfine and flavor splittings
of the heavy-light mesons in one-loop diagrams. 
We follow the argument of Ref.~\cite{FermilabMILC_Dec2011} and 
briefly describe how one can adjust \eq{1p1p1_pq_fD} to include these splittings. 
In \eq{1p1p1_pq_fD}, the contributions proportional to $g_\pi^2$ come from diagrams with
internal $D^*$ propagators, and the contributions with no factor of  $g_\pi^2$ come from
diagrams with light-meson (``pion'') tadpoles.  Thus we must only adjust the former contributions.
The splittings in diagrams with internal $D^*$ propagators depend on whether the pion line is connected,
which results in the term with the sum over $\mathscr{S}$ in \eq{1p1p1_pq_fD}, or disconnected,
which results in the terms with the factors of the residue function $R$ in \eq{1p1p1_pq_fD}.
(See Fig.~5 in Ref.~\cite{Aubin:StagHL2006} for the structure of the quark flow in these diagrams.) 
In the disconnected case, the valence $x$ quark in the external $D_{x\Xi}$
flows into the pion propagator and then returns the way it came (a ``hairpin''
diagram) and enters the $D^*$ propagator.  Thus the internal $D^*$ always has the same
flavor as the external $D_{x\Xi}$, so there is no flavor splitting between the two,
only a hyperfine splitting.  In the connected case, the $D^*$ in the loop has the flavor of
the virtual sea quark loop (which we labeled by $\mathscr{S}$ in \eq{1p1p1_pq_fD}), 
so there is flavor splitting with the external $D_{x\Xi}$, in addition
to the  hyperfine splitting.

We let  $\Delta^*$ be the lowest-order hyperfine splitting, and $\delta_{\mathscr{S}x}$  be the
flavor splitting between a heavy-light meson with  light quark of flavor $\mathscr{S}$ and one
of flavor $x$. At lowest order, $\delta_{\mathscr{S}x}$ is  proportional to the quark-mass
difference,
which can be written in terms of the parameter $\lambda_1$ in \eq{L2m}:
\begin{equation}\label{eq:deltaeq}
    \delta_{\mathscr{S}x} \cong 2 \lambda_1 (m_\mathscr{S}-m_x)  \cong
    \frac{\lambda_1}{\mu} (m^2_{\mathscr{S}\mathscr{S},\xi_5} - m^2_{xx,\xi_5}) ,
\end{equation}
where the final expression expresses the result in terms of pion masses.

Since the mass of the external $D$ is removed in HQET, 
the mass shell is at $k = 0$. 
When there is no splitting, the internal $D^*$ has its pole at the same place, which makes
the integrals simple and gives rise to the chiral log function
$\ell(m^2)$.
In the presence of a splitting $\Delta$ between the internal $D^*$ and the external $D$,
the integrals involve the more complicated function
\begin{equation}
    J(m,\Delta) = (m^2-2\Delta^2)\log(m^2/\Lambda^2) +2\Delta^2 -4\Delta^2 F(m/\Delta) . \label{eq:Jdef}
\end{equation}
Here the function $F$ is \cite{Stewart:1998,Becirevic:2003}
\begin{equation}
F(1/x) = 
    \begin{cases} 
     -\frac{\sqrt{1-x^2}}{x_{\phantom{g}}}\left[\frac{\pi}{2} - \tan^{-1}\frac{x}
    {\sqrt{1-x^2}}\right], & \text{if $ |x|\le 1$,} \\
    \frac{\sqrt{x^2-1}}{x}\ln(x + \sqrt{x^2-1}), & \text {if $ |x|\ge 1$ \ .}
    \end{cases}
    \label{eq:Fdef}
\end{equation}

We may now generalize \eq{1p1p1_pq_fD} to include splittings.
We simply replace
\begin{equation}
    \ell(m^2) \to  J(m,\Delta)
    \label{eq:replacement}
\end{equation}
in the terms proportional to $g_\pi^2$, taking care to include the flavor splittings 
($\Delta=\Delta^* + \delta_{\mathscr{S}x}$) for terms from connected-pion diagrams, and to omit 
the flavor splittings ($\Delta=\Delta^*$) for terms from disconnected-pion diagrams.
The result for the  leptonic decay constant is then
\begin{eqnarray}\label{eq:1p1p1_pq_fD_withsplittings}
\frac{ f_{D_{x\Xi}} \sqrt{ M_{D_{x\Xi}} } }{\kappa}
      & = &        1 + \frac{1}{16\pi^2f^2} \frac{1}{2}
      \Biggl\{-\frac{1}{16}\sum_{\mathscr{S},\Xi'} \ell(m_{\mathscr{S}x,\Xi'}^2)
	  \nonumber \\*&&{}- \frac{1}{3}
	  \sum_{j\in \cM_I^{(3,x)}} 
	  \frac{\partial}{\partial m^2_{X,I}}\left[ 
		R^{[3,3]}_{j}(
	      \cM_I^{(3,x)};  \mu^{(3)}_I) \ell(m_{j}^2) \right]
	      \nonumber \\*&&{} 
	      -   \biggl( a^2\delta'_V \sum_{j\in \cM_V^{(4,x)}}
	      \frac{\partial}{\partial m^2_{X,V}}\left[ 
		R^{[4,3]}_{j}( \cM_V^{(4,x)}; \mu^{(3)}_V)
	      \ell(m_{j}^2)\right]
		  + [V\to A]\biggr) \nonumber \\*&&{}
	    -3g_\pi^2\frac{1}{16}\sum_{\mathscr{S},\Xi'} J(m_{\mathscr{S}x,\Xi'},\Delta^*+\delta_{\mathscr{S}x})
	    \nonumber \\*&&{}- g_\pi^2
	  \sum_{j\in \cM_I^{(3,x)}} 
	  \frac{\partial}{\partial m^2_{X,I}}\left[ 
		R^{[3,3]}_{j}(
	      \cM_I^{(3,x)};  \mu^{(3)}_I) J(m_{j},\Delta^*) \right]
	      \nonumber \\*&&{} 
	      \hspace{0cm} -3g_\pi^2 \biggl( a^2\delta'_V \sum_{j\in \cM_V^{(4,x)}}
	      \frac{\partial}{\partial m^2_{X,V}}\left[ 
		R^{[4,3]}_{j}( \cM_V^{(4,x)}; \mu^{(3)}_V)
	      J(m_{j},\Delta^*)\right]
		  + [V\to A]\biggr)  
	    \Biggr\} \nonumber \\*&&{}+
      c_s (m_u + m_d + m_s) + c_v m_x + c_{a,\Xi} a^2 \ .
\end{eqnarray}
We can also include the finite-volume effects for a spatial volume~$L^3$ into \eq{1p1p1_pq_fD_withsplittings}.  
Following Ref.~\cite{FermilabMILC_Dec2011}, we replace
\begin{eqnarray}
      \ell( m^2)  &\to & \ell( m^2)  + m^2 \delta_1(mL) ,\\
      J(m,\Delta) &\to & J(m,\Delta) + \delta J(m,\Delta,L) ,
\end{eqnarray}
where
\begin{equation}
\delta J(m,\Delta,L)  = \frac{m^2}{3}\delta_1(mL) - 16\pi^2\left[\frac{2\Delta}{3}J_{FV}(m,\Delta,L)
+\frac{\Delta^2-m^2}{3} K_{FV}(m,\Delta,L)\right] \ ,
\end{equation}
with
\begin{equation}
    K_{FV}(m,\Delta,L) \equiv \frac{\partial}{\partial \Delta} J_{FV}(m,\Delta,L) ,
\end{equation}
and with $\delta_1(mL)$ and $J_{FV}(m,\Delta,L)$ defined in Refs.~\cite{Aubin:StagHL2007,Arndt:2004}.
 
Reference \cite{FermilabMILC_Dec2011} also discusses the extent to 
which including the splittings as in \eq{1p1p1_pq_fD_withsplittings}, and not other possible $1/m_Q$
effects, is a systematic improvement on \eq{1p1p1_pq_fD}. In that discussion the power counting introduced 
by Boyd and Grinstein \cite{BoydGrinstein} is applied, which assumes
\begin{equation}
    \frac{\Delta^2,\; \Delta m,\; m^2}{m_Q} \ll \Delta \sim m \ ,
    \label{eq:power-counting-BoydGrinstein}
\end{equation}
where $\Delta$ is a generic splitting ($\Delta^*$ or $\delta_{\mathscr{S}x}$ or a linear combination of the two), 
$m$ is a generic light pseudoscalar {\it meson}\/ mass, and $m_Q$ is the heavy quark mass. In the lattice simulations of
Ref.~\cite{FermilabMILC_Dec2011}, the lowest pion masses were about half the physical kaon mass, 
and the power counting
of Ref.~\cite{BoydGrinstein} was only marginally applicable to the data.   However, for simulations
on the HISQ ensembles generated by the MILC Collaboration \cite{Bazavov:2012xda,Bazavov:2010ru}, the lowest pion
masses are physical, and the assumptions of the
Boyd-Grinstein power counting are well satisfied.  Furthermore, including the splittings with such data
is not optional: for $D$ mesons the hyperfine splitting $\Delta^*=142.1\;$MeV, and the
flavor splitting $\delta_{sd}=98.9\;$MeV, clearly non-negligible compared to the physical pion mass.

Since we have included hyperfine and flavor splittings, which are empirically large even though they
are formally of order $1/m_Q$, it is
important to consider whether splittings coming from taste violations should also
be included in the heavy-light propagators at one loop. 
As discussed in the introduction, taste splittings in squared meson masses are roughly constant as
the masses increase from pions to $D$ mesons, which means that taste splittings in the heavy-light
masses themselves are quite small, $\sim\! 11\;$MeV at $a\approx0.12\;$fm for the HISQ action. 
The taste splittings are indeed higher order compared to the physical hyperfine and flavor splittings.
We note that the taste-violating Lagrangian terms in \eqs{L2A2}{L2B2} also lead to $\cO(a^2)$ contributions to hyperfine splittings. Those effects have not been measured in lattice simulations, but we think it is reasonable to assume they are comparable in size to the taste splittings since in most cases the same operators produce both
effects.

There is also the question of whether other $1/m_Q$ continuum effects should be included along with the
hyperfine and flavor splittings. As discussed in Ref.~\cite{FermilabMILC_Dec2011},  
such terms only change the overall normalization of the result for the  quantity $ \delta\! f_{D_{x}} $ in \eq{fD}
by relatively small amount, of order $\Lambda_{QCD}/m_Q$.  Since in any case the value of $f$ in \eq{fD} may
be considered uncertain by as much as $20\%$ (the difference between $f_\pi$ and $f_K$), these additional $1/m_Q$ terms have no practical implications for our results.

\section{Conclusions\label{sec:conclusions}}

We have generalized the chiral Lagrangian for heavy-light mesons to the case where
both heavy and light quarks have the staggered action.  A fundamental assumption
of our work is that lattice spacings is sufficiently small, or 
the heavy-quark action is sufficiently improved, that we may treat $a m_Q$ as a small parameter, where
$m_Q$ is the heavy quark mass.
This is the same assumption required in order to describe heavy quarks with the HISQ staggered
action in simulations.

The heavy-light part of the LO staggered chiral Lagrangian we obtain is 
identical to that in the continuum, except for extra taste degrees of freedom of the light and
heavy quarks.
In contrast with the light-light part of the chiral Lagrangian, which includes taste splittings at LO,
the heavy-light part of the LO chiral Lagrangian is taste-invariant, with three key symmetries:
heavy quark spin symmetry,  chiral symmetry of the light quarks (including taste and flavor
symmetries), and $SU(4)$ taste symmetry of the heavy quarks.
Complications arise at NLO, where these symmetries of the heavy-light Lagrangian may be broken
by lattice artifacts, as well as by light-quark mass terms.  Those NLO contributions that
arise from terms in the Symanzik effective theory composed
exclusively of light quarks may be taken over directly from \rcite{Aubin:StagHL2006}.
In doing so, we have corrected some minor errors in that reference, which do not affect any existing calculations within that framework.  Terms in the Symanzik effective theory with heavy staggered quarks are new. We have derived their consequences for the NLO heavy-light
Lagrangian, as well as the left-handed current, in some detail. In some cases, though, we have not 
attempted to find the complete set of possible terms, and have contented ourselves with simply
listing sufficient numbers of terms relevant to foreseeable practical applications.  

We have then applied our Lagrangian to calculate, through NLO,  the taste splitting of heavy-light
mesons and the heavy-light leptonic decay constant.  In both these cases, we are able to 
prove that the one-loop diagrams are taste invariant, despite the fact that they contain pion
propagators that break taste symmetry.  This means that taste violations
in these quantities at NLO come exclusively from analytic terms, which arise from
the NLO Lagrangian and current.
Using our results for the mass splittings, and making assumptions about the dominant
operators based on experience with light-light quantities, we find that we can qualitatively
understand  
the pattern of splittings seen in heavy-light HISQ data.

For the decay constant, the NLO taste violations produce a single analytic term that depends on taste of the meson, the term $c_{a,\Xi}a^2$ in \eqs{1p1p1_pq_fD}{1p1p1_pq_fD_withsplittings}. The
one-loop diagrams give rise to the same chiral logarithms derived in \rcite{Aubin:StagHL2006}, because
in both cases they are taste invariant. 
Following \rcite{FermilabMILC_Dec2011}, we include the modifications of these chiral logarithms
due to heavy-light hyperfine and flavor splittings, which are comparable in size to the physical
pion mass, and therefore important for describing modern simulations in which the light quark
 masses are physical or close to physical.  The resulting chiral form is being used to fit HISQ data for 
 decay constants of the $D$ system \cite{work-in-progress}.  Although such
fits may be bypassed for data at physical quark masses \cite{Bazavov:2012dg}, the  chiral fits allow one 
to include data at unphysical quark masses, and thereby one can hope to obtain smaller statistical errors and better control over 
continuum extrapolation errors. The work in progress indicates that these hopes are realized in practice.

\bigskip
\bigskip
\centerline{\bf ACKNOWLEDGMENTS}
\bigskip
We thank X.\ Du, M.\ Lightman and A.\ Kronfeld for helpful discussions.
We are grateful to our colleagues in MILC for the use of data on HISQ 
splittings.

\appendix
\section{\label{sec:tensor} Reduction to Irreducible Tensors}

In \secref{NLOa2type2}, we need to reduce a 3-index tensor to irreducible
Lorentz representations in order to find the type-B2 chiral form for the current.  
The reduction is done explicitly here.  For present
convenience we work in Euclidean space and use Euclidean rotational symmetry (plus parity)
instead of Lorentz symmetry, so we do not have to worry about upper and lower indices.

Consider a tensor $X^{\alpha\beta\rho}$, which may be taken to be traceless on the
second two indices $X^{\alpha\lambda\lambda}=0$, where sum over $\lambda$ is implied.%
\footnote{In what follows $\lambda$ will be used as a summation index, and sum over it
is always implied when it appears twice.  However, all other indices are not summed over, even
when they appear more than once.}
The tracelessness may be assumed because
the trace term will simply reproduce type-A contributions, as in the discussion of 
$\cL^{B2}_{2,a^2}$. 
In addition, we may just consider the reduction
of the part of $X$ that is symmetric on the second two indices,
\begin{equation}
Y^{\alpha\beta\rho} \equiv \frac{1}{2}\left(X^{\alpha\beta\rho}+X^{\alpha\rho\beta}\right)
\eqn{Y}
\end{equation}
since we will ultimately be interested 
in writing only the element $X^{\mu\nu\nu}$ in terms of irreducible tensors, and
the antisymmetric part will not contribute.

The tensor $Y$ transforms as the product of a vector (on the first index) and a traceless
symmetric tensor (on the second and third indices).  To see what representations appear,
we use the fact that $SO(4)=SU(2)\times SU(2)$ to denote irreducible tensors by their
spin under the two $SU(2)$ factors.  A vector is the $(\half,\half)$ representation, while
a traceless, symmetric two-index tensor is the $(1,1)$ representation.  The product
thus contains $(\threehalves,\threehalves)$, $(\threehalves,\half)\oplus(\half,\threehalves)$, 
and $(\half,\half)$, where parity interchanges the two $SU(2)$ factors, 
making a single representation out of the second component.  The highest representation must be symmetric, 
and $(\threehalves,\threehalves)$ corresponds to a completely symmetric, three index tensor 
$S^{\alpha\beta\rho}$, which is traceless
on any pair of indices: $S^{\alpha\lambda\lambda}=
S^{\lambda\alpha\lambda}=S^{\lambda\lambda\alpha}=0$.  The $(\threehalves,\half)
\oplus(\half,\threehalves)$ is a traceless
three index tensor $A^{\alpha\beta\rho}$ of mixed symmetry, antisymmetric on the first two
indices (say).  The $(\half,\half)$ is a vector $W^\rho$, formed
from only the nonvanishing trace of $Y$:
\begin{equation}
W^\rho = Y^{\lambda\lambda\rho}=  Y^{\lambda\rho\lambda} \eqn{W} \ .
\end{equation}
Constructing $S^{\alpha\beta\rho}$ and $A^{\alpha\beta\rho}$, we have
\begin{eqnarray}
S^{\alpha\beta\rho} &=& \frac{1}{3}
\left(Y^{\alpha\beta\rho}+Y^{\beta\rho\alpha}
+ Y^{\rho\alpha\beta}\right) 
-\frac{1}{9}
\left(\delta^{\beta\rho}\;Y^{\lambda\lambda\alpha} +\delta^{\alpha\rho}\;  
Y^{\lambda\lambda\beta} + \delta^{\alpha\beta}\;  Y^{\lambda\lambda\rho}\right)\ , \eqn{S}\\
A^{\alpha\beta\rho} &=& \frac{1}{2}
\left(Y^{\alpha\beta\rho}-Y^{\beta\alpha\rho} \right) 
-\frac{1}{6}
\left(\delta^{\alpha\rho}\;  Y^{\lambda\lambda\beta}-\delta^{\beta\rho}\;Y^{\lambda\lambda\alpha} 
\right)\ . \eqn{A}
\end{eqnarray}
From the $SU(2)\times SU(2)$ quantum numbers, $S$ and $A$ should each be 16-dimensional.
Checking this for $S$ is straightforward; for $A$, the following identity is helpful:
\begin{equation}
A^{\alpha\beta\rho}+A^{\beta\rho\alpha}
+ A^{\rho\alpha\beta} = 0\ . \eqn{Aidentity}
\end{equation}

Solving \eqsthru{W}{A} for $Y^{\alpha\beta\rho}$ gives the reduction
\begin{equation}
Y^{\alpha\beta\rho} = S^{\alpha\beta\rho} + \frac{2}{3}\left(A^{\alpha\beta\rho}-
A^{\rho\alpha\beta}\right) +\frac{1}{9}
\left(2\,\delta^{\alpha\rho}\;W^{\beta} +2\,\delta^{\alpha\beta}\;
W^{\rho} - \delta^{\beta\rho}\;  W^{\alpha}\right)\ . \eqn{Yred}
\end{equation}
The particular case of interest is the reduction of $X^{\mu\nu\nu}$. From \eqs{Y}{Yred}, 
we have
\begin{equation}
X^{\mu\nu\nu} = S^{\mu\nu\nu} + \frac{4}{3}A^{\mu\nu\nu}
+\frac{1}{9}
\left(4\,\delta^{\mu\nu}\;W^{\nu} -  W^{\mu}\right)\ . \eqn{Xred}
\end{equation}

\end{document}